\documentclass[letterpaper,11pt,final]{elsarticle}

\usepackage[utf8]{inputenc} 
\usepackage[T1]{fontenc}    
\usepackage[semibold,tabular]{sourceserifpro}
\usepackage[charter,scaled=1.08]{newtxmath}
\usepackage[scaled=1.1]{zlmtt}
\usepackage{multirow}
\usepackage[english]{babel}

\usepackage[frozencache,cachedir=cache]{minted}
\usepackage{xcolor}
\usepackage{xspace}
\usepackage{bbm}
\usepackage[letterspace=50]{microtype}
\usepackage{ragged2e}
\usepackage{relsize}
\usepackage{acro}
\acsetup{
  format/short = {\lsstyle\scshape}
}
\makeatletter
\let\@afterindenttrue\@afterindentfalse
\makeatother

\DeclareAcronym{ABF}{short = abf, long = adaptive biasing force}
\DeclareAcronym{ASE}{short = ase, long = Atomic Simulation Environment, first-style = short}
\DeclareAcronym{ANN}{short = ann, long = artificial neural networks sampling}
\DeclareAcronym{BCC}{short = bcc, long = body-centered cubic}
\DeclareAcronym{CFF}{short = cff, long = combined force frequency sampling}
\DeclareAcronym{CV}{short = cv, long = collective variable, pdfstring = CV}
\DeclareAcronym{CPU}{short = cpu, long = central processing unit, first-style = short, pdfstring = CPU}
\DeclareAcronym{FFS}{short = ffs, long = forward flux sampling}
\DeclareAcronym{FUNN}{short = funn, long = force-biasing using neural networks}
\DeclareAcronym{JSON}{short = json, long = JavaScript Object Notation, first-style = short}
\DeclareAcronym{WTMD}{short = wtmd, long = well-tempered  metadynamics}
\DeclareAcronym{FES}{short = fes, long = free energy surface, short-plural = {}}
\DeclareAcronym{GAP}{short = gap, long = Gaussian Approximation Potential}
\DeclareAcronym{GPU}{short = gpu, long = graphics processing unit, first-style = short, pdfstring = GPU}
\DeclareAcronym{TPU}{short = tpu, long = tensor processing unit, first-style = short}
\DeclareAcronym{DFT}{short = dft, long = density functional theory}
\DeclareAcronym{LDA}{short = lda, long = local density approximation, first-style = short}
\DeclareAcronym{MGI}{short = mgi, long = materials genome initiative}
\DeclareAcronym{SSAGES}{
  short = ssages,
  long = Software Suite for Advanced General Ensemble Simulations
}
\DeclareAcronym{HCG}{short = hcg, long = highly coarse-grained}     
\DeclareAcronym{HPC}{short = hpc, long = high performance computing}
\DeclareAcronym{LC}{short = lc, long = liquid crystal}
\DeclareAcronym{LJ}{short = lj, long = Lennard--Jones}
\DeclareAcronym{MFEP}{short = mfep, long = mean free energy pathway}
\DeclareAcronym{MPI}{short = mpi, long = message passing interface, first-style = short}
\DeclareAcronym{MD}{short = md, long = molecular dynamics}
\DeclareAcronym{ML}{short = ml, long = machine learning}
\DeclareAcronym{NN}{short = nn, long = neural network}
\DeclareAcronym{GNN}{short = gnn, long = Graph neural network}
\DeclareAcronym{DPD}{short = dpd, long = dissipative particle dynamics}
\DeclareAcronym{PME}{short = pme, long = particle mesh Ewald}
\DeclareAcronym{RCC}{short = rcc, long = Research Computing Center}
\DeclareAcronym{TPS}{short = tps, long = time steps per second}
\DeclareAcronym{DOF}{short = dof, long = degrees of freedom}
\DeclareAcronym{PLUMED}{short = plumed, long = PLugin for MolEcular Dynamics, first-style = short}
\DeclareAcronym{SMILES}{short = smiles, long = simplified molecular-input line-entry system, first-style = short}
\DeclareAcronym{5CB}{short = 5CB, long = 4-cyano-4'-pentylbiphenyl, short-format = {\lsstyle\scshape\MakeLowercase}}
\DeclareAcronym{SDS}{short = sds, long = sodium lauryl sulfate}
\DeclareAcronym{ADP}{short = adp, long = alanine dipeptide}
\DeclareAcronym{ns/day}{short = ns/day, long = nano-seconds per day, format = {}}

\newcommand{\version}[1]{v\kern 0.1em\textsc{#1}}
\newcommand{\caps}[1]{\textls[50]{\textsc{\MakeLowercase{#1}}}}
\newcommand*{\emdash}{\;\rule[.5ex]{.5em}{.5pt}\;}

\newcommand*{\PySAGES}{\texorpdfstring{\textls[50]{\textsc{p}y\textsc{sages}}\xspace}{PySAGES}}
\newcommand*{\PPySAGES}{\texorpdfstring{\textls[50]{Py\textsc{sages}}\xspace}{PySAGES}}
\newcommand*{\JAX}{\caps{JAX}\xspace}
\newcommand*{\JAXMD}{\JAX~\caps{MD}\xspace}
\newcommand*{\HOOMD}{\caps{hoomd-}blue\xspace}
\newcommand*{\LAMMPS}{\textls[60]{\textsmaller{LAMMPS}}\xspace}
\newcommand*{\OpenMM}{\textls[25]{Open}\caps{MM}\xspace}
\newcommand*{\DLPack}{\caps{DLP}ack\xspace}
\newcommand*{\DuCK}{\textls[50]{\textsc{d}u\textsc{ck}}\xspace}
\newcommand*{\DeepMD}{\textls[25]{Deep}\caps{MD}\xspace}
\newcommand*{\BigSMILES}{\textls[40]{Big\kern0.05em\textsc{smiles}}\xspace}
\newcommand*{\CuPy}{\textls[50]{\textsc{c}u\textsc{p}y}\xspace}

\definecolor{pmeblue}{HTML}{007ba0}
\definecolor{pmered}{HTML}{8a0021}
\definecolor{pmegrey}{HTML}{bdb9b5}
\usepackage[colorlinks]{hyperref}
\AtBeginDocument{%
  \hypersetup{
    breaklinks,
    citecolor={pmeblue!60!black},
    linkcolor={pmeblue!60!black},
    urlcolor={pmeblue!60!black}
  }%
}
\graphicspath{{figures/}}

\makeatletter
\@ifpackageloaded{babel}{
  \addto\extrasenglish{%
    \renewcommand{\appendixautorefname}[1]{}%
  }%
}{\relax}
\makeatother

\begin{document}

\begin{frontmatter}



\title{\texorpdfstring{\textls[80]{PySAGES}}{PySAGES}: flexible, advanced sampling methods accelerated with \acp{GPU}}


\author[pme]{Pablo F. {Zubieta~Rico}}
\author[pme]{Ludwig Schneider}
\author[pme]{Gustavo R. Pérez-Lemus}
\author[pme]{Riccardo Alessandri}
\author[pme]{Siva Dasetty}
\author[pme]{Cintia A. Menéndez}
\author[pme]{Yiheng Wu}
\author[pme]{Yezhi Jin}
\author[pme]{Yinan Xu}
\author[rcc]{Trung D. Nguyen}
\author[rcc]{John A. Parker}
\author[pme]{Andrew L. Ferguson}
\author[ndu]{Jonathan K. Whitmer}
\author[pme]{Juan J. {de~Pablo}}

\affiliation[pme]{
  organization={Pritzker School of Molecular Engineering, The University of Chicago},
  addressline={5640 South Ellis Avenue}, 
  city={Chicago},
  state={IL},
  postcode={60637},
  country={USA}
}

\affiliation[rcc]{
  organization={Research Computing Center, The University of Chicago},
  addressline={6054 S. Drexel Avenue}, 
  city={Chicago},
  state={IL},
  postcode={60637},
  country={USA}
}

\affiliation[ndu]{
  organization={Department of Chemical and Biomolecular Engineering, University of Notre Dame},
  addressline={250 Nieuwland Hall}, 
  city={Notre Dame},
  state={IN},
  postcode={46556},
  country={USA}
}

\begin{abstract}
Molecular simulations are an important tool for research in physics, chemistry, and biology. The capabilities of simulations can be greatly expanded by providing access to advanced sampling methods and techniques that permit calculation of the relevant underlying free energy landscapes. In this sense, software that can be seamlessly adapted to a broad range of complex systems is essential. Building on past efforts to provide open-source community supported software for advanced sampling, we introduce \PySAGES, a Python implementation of the \ac{SSAGES} that provides full \ac{GPU} support for massively parallel applications of enhanced sampling methods such as adaptive biasing forces, harmonic bias, or forward flux sampling in the context of molecular dynamics simulations. By providing an intuitive interface that facilitates the management of a system's configuration, the inclusion of new collective variables, and the implementation of sophisticated free energy-based sampling methods, the \PySAGES library serves as a general platform for the development and implementation of emerging simulation techniques. The capabilities, core features, and computational performance of this new tool are demonstrated with clear and concise examples pertaining to different classes of molecular systems.
We anticipate that \PySAGES will provide the scientific community with a robust and easily accessible platform to accelerate simulations, improve sampling, and enable facile estimation of free energies for a wide range of materials and processes.
\end{abstract}




\end{frontmatter}


\section{Introduction}
\label{sec:introduction}


Molecular simulations are extensively used in a wide range of science and engineering disciplines\;\cite{nobelprize2013chemistry}. As their use has grown for the discovery of new phenomena and the interpretation of sophisticated experimental measurements, so has the complexity of the systems that are considered.
Classical atomistic \ac{MD} simulations are generally limited to microsecond time scales and length scales of tens of nanometers. For systems that are characterized by rugged free energy landscapes, such time scales can be inadequate to ensure sufficient sampling of the relevant phase space, and advanced methods must therefore be adopted to overcome free energy barriers. 
In that regard, it is useful and increasingly common to identify properly chosen \acp{CV},
which are generally differentiable functions of the atomic coordinates of the system; then, biases can be applied to explore the space defined by such \acp{CV}, thereby overcoming barriers and enhancing sampling of the thermally accessible phase space.

The rapid growth of hardware accelerators such as \acp{GPU} or \acp{TPU}, or specialized hardware designed for fast \ac{MD} computations\;\cite{shaw2008anton,shaw2014anton}, has provided researchers with increased opportunities to perform longer simulations of larger systems. \Acp{GPU}, in particular, provide a widely accessible option for fast simulations, and several software packages, such as \HOOMD\;\cite{HOOMD-blue}, \OpenMM\;\cite{OpenMM}, \JAXMD\;\cite{jaxmd2020,schoenholz2021jax}, \LAMMPS\;\cite{LAMMPS}, and Gromacs\;\cite{abraham2015gromacs}, are now available for \ac{MD} simulations on such devices.

As mentioned above, enhanced sampling methods seek to surmount the high energy barriers that separate multiple metastable states in a system, while facilitating the calculation of relevant thermodynamic quantities as functions of different \acp{CV} such as \acp{FES}. 
Several libraries, such as \ac{PLUMED}\;\cite{tribello2014plumed}, Colvars\;\cite{fiorin2013using},
and our own \ac{SSAGES} package\;\cite{sidky2018ssages}, provide out-of-the-box solutions for performing enhanced sampling \ac{MD} simulations.

Among the various enhanced sampling methods available in the literature, some of the most recently devised schemes rely on \ac{ML} strategies to approximate free energy surfaces and their gradients (generalized
forces)\;\cite{2018HSidky-JCP-ANN,2018AZGuo-JCP-FUNN,2020ESevgen-JCTC-CFF, wang2022efficient}.
Similarly, algorithms for identifying meaningful \acp{CV} that
correlate with high variance or slow \acp{DOF} are based on deep neural networks\;\cite{schwantes2013improvements,chen2018molecular,mardt2018vampnets,chen2019capabilities,chen2019nonlinear,sidky2020molecular}.
These advances serve to highlight the need for seamless integration of \ac{ML} frameworks with existing \ac{MD} software libraries.

To date, there are no solutions that combine enhanced sampling techniques, hardware
acceleration, and \ac{ML} frameworks to facilitate enhanced-sampling \ac{MD}
simulations on \acp{GPU}.
While some \ac{MD} libraries that support \acp{GPU} provide
access to a limited set of enhanced sampling methods\;\cite{OpenMM,abraham2015gromacs,lee2018gpu,phillips2020scalable,kobayashi2017genesis},
there are currently no packages that
enable users to take advantage of all of these features within the same platform and in the same backend-agnostic fashion that tools such as \ac{PLUMED} and \ac{SSAGES}
have provided for \ac{CPU}-based \ac{MD} simulations.


Here we present \PySAGES, a Python Suite for Advanced General Ensemble
Simulations.
It is a free, open-source software package written in
Python and based on \JAX that follows the design ideas of \ac{SSAGES} and enables users to easily perform enhanced-sampling \ac{MD} 
simulations on \acp{CPU}, \acp{GPU}, and \acp{TPU}.
\PPySAGES can currently be coupled with \HOOMD,
\OpenMM, \JAXMD and {\ac{ASE}}\emdash and by extension from the latter to \caps{CP2K}, Quantum \caps{ESPRESSO}, \caps{VASP} and Gaussian, among others. 
At this time, \PySAGES offers the following
enhanced sampling methods: Umbrella Sampling, Metadynamics,
Well-tempered Metadynamics, Forward Flux Sampling, String Method,
Adaptive Biasing Force, Artificial neural network sampling, Adaptive
Biasing Force using neural networks, Combined Force Frequency, and
Spectral Adaptive Biasing Force.
\PPySAGES also includes some of the most
commonly used \acp{CV} and, importantly, defining new ones is relatively simple, as
long as they can be expressed in terms of the NumPy\;\cite{harris2020array}
interface provided by \JAX. All \acp{CV} can be
automatically differentiated through \JAX functional transforms.
\PPySAGES is highly modular, thereby allowing for the easy implementation of new methods as they emerge, even as part of a user-facing script.

In the following sections, we provide a general overview of the design and implementation of \PySAGES , and present a series of examples to showcase its flexibility for addressing research problems in different application areas. We also discuss its performance in \acp{GPU} and present a few perspectives on how to grow and improve the package to cover more research use cases through future development, as well as community involvement and contributions.

\section{Implementation}
\label{sec:implementation}

\begin{figure}
  \centering
  \includegraphics[width=0.875\textwidth]{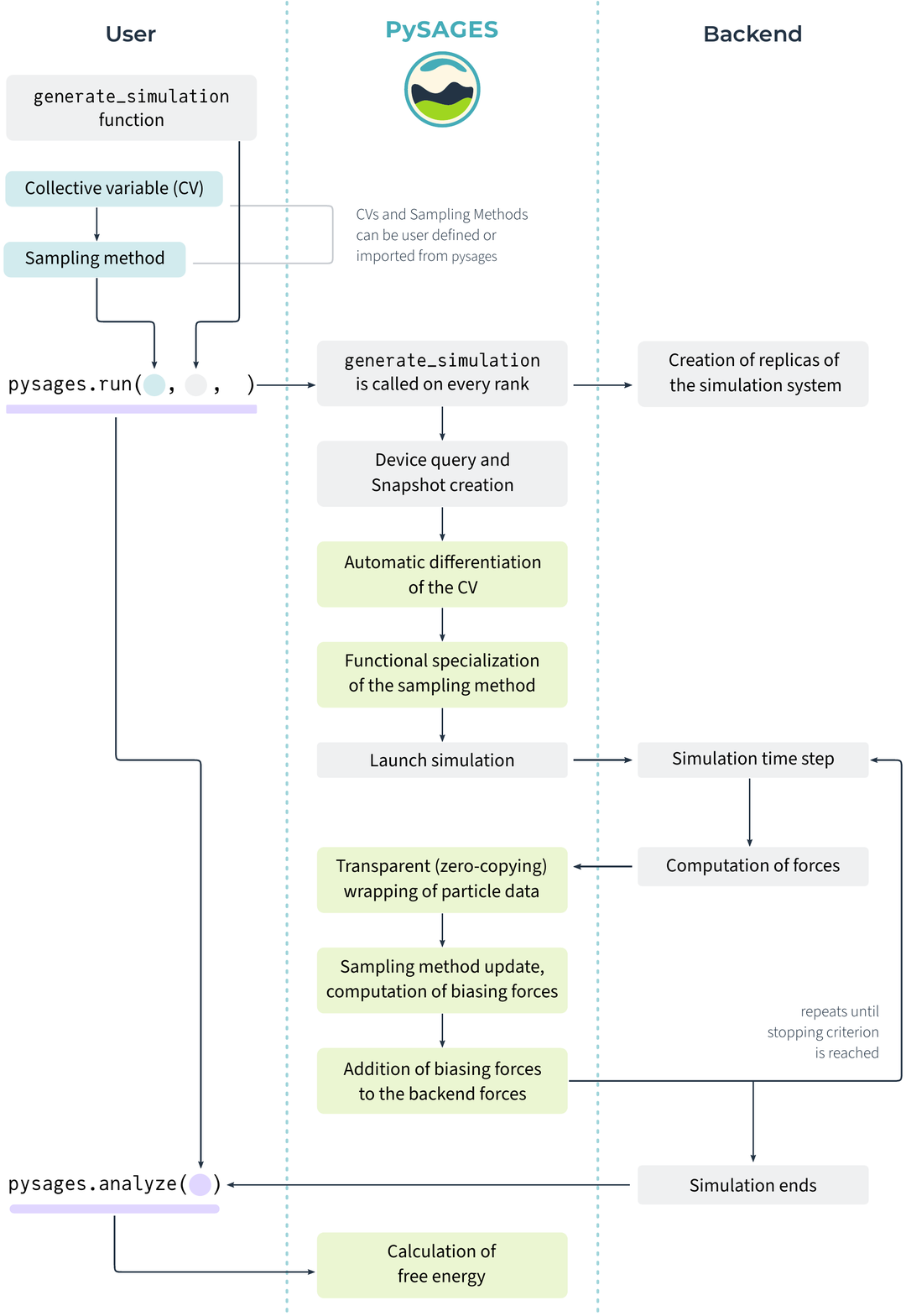}
  \caption{
    The \PySAGES simulation flowchart. For a simulation, a user sets up a script that declares the \ac{CV} and sampling methods to be used.
  }
  \label{fig:flowchart}
\end{figure}

We begin by briefly outlining the core components of
\PySAGES, how they function together, and how communication with
each backend allows \PySAGES to bias a simulation during runtime.
A summary of the execution workflow of \PySAGES
along with a mapping of the user interface with the main stages
of the simulation and the interaction with the backends, is
illustrated in \autoref{fig:flowchart}.

\begin{figure}
  \centering
  \includegraphics[width=\textwidth]{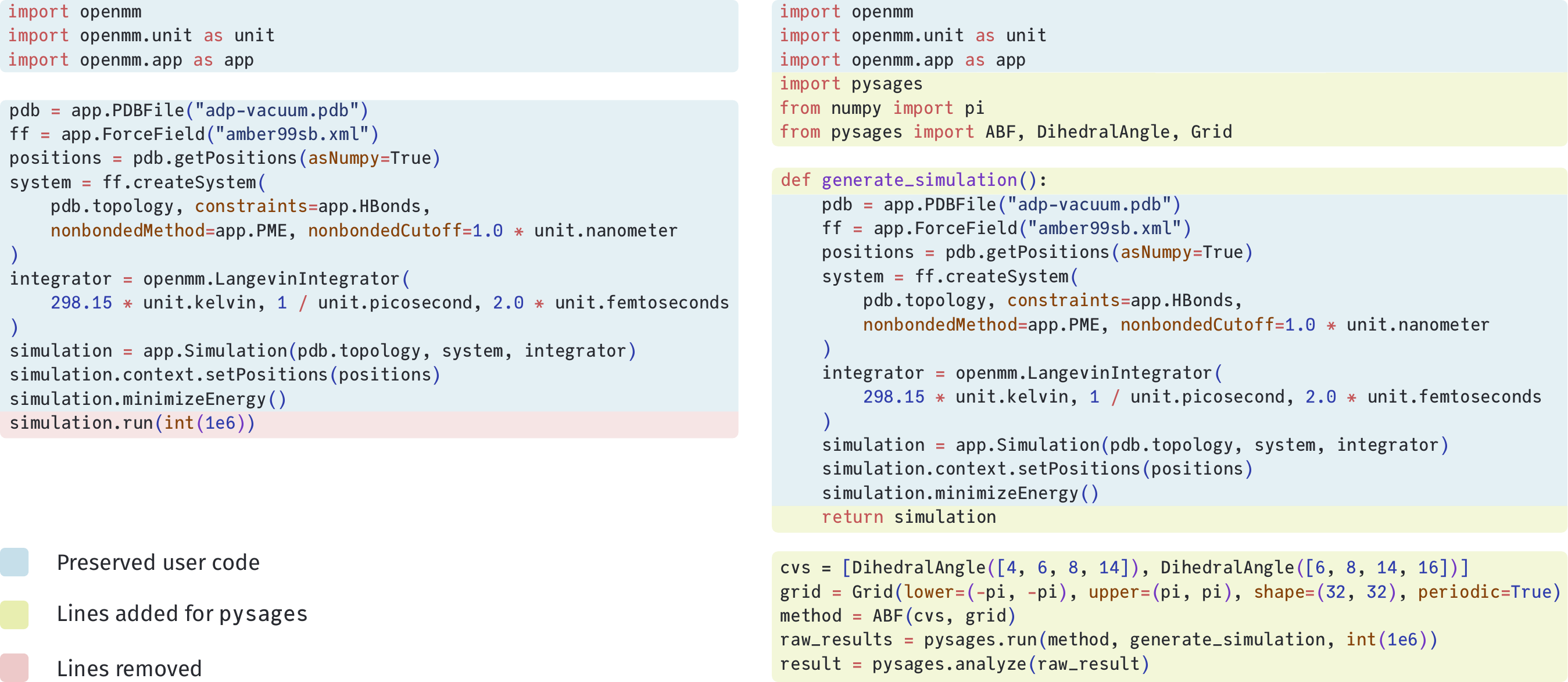}
  \caption{
    Example of how to use the Python interface for \PySAGES. It is easy to extend existing \ac{MD} scripts with \PySAGES to perform enhanced-sampling \ac{MD}, with minimal changes to the code.
    In general, the only requirement is for the user to wrap the code that
    defines the simulation system into a \emph{simulation generator} function.
  }
  \label{fig:code-comparison}
\end{figure}

To provide a uniform user interface while minimizing disruption to preexisting workflows, \PySAGES only requires the user to wrap their
traditional backend scripting code into \emph{simulation generator} functions.
This approach accommodates the heterogeneity of Python interfaces across the different simulation backends supported by \PySAGES.
An example of a simulation generator function and how a traditional
\OpenMM script can be modified to perform an enhanced-sampling \ac{MD} simulation is depicted in \autoref{fig:code-comparison}.

At the start of a simulation, the simulation generator function is
called to instantiate as many replicas of the simulation as needed.
Then, for each replica, \PySAGES queries the particle information
and the device that the backend will be using. In addition,
during this initial stage \PySAGES also performs automatic
differentiation of the collective variables via \JAX's \texttt{grad}
transform\emdash required to estimate the biasing forces,
and generates specialized initialization and updating routines
for the user-declared sampling method.

Like \ac{SSAGES}, \PySAGES wraps the simulation
information into an object called a \texttt{Snapshot}. This object
exposes the most important simulation information, such as particle
positions, velocities, and forces in a backend- and device-agnostic format. To achieve this, \PySAGES uses
\DLPack\;\cite{dlpack}\emdash for \caps{C++} based \ac{MD} libraries\emdash%
to directly access the contents of the backend-allocated buffers
for the different particle properties without creating data
copies whenever possible.

Once the setup of both the simulation and sampling method is completed,
\PySAGES hands control back to the backend, which will run for a
given number of time steps or until some other stopping criteria is
reached. In order to exchange information back and forth, \PySAGES
adds a force-like object or function to the backend which gets
called as part of the time integration routine. Here, the
sampling method state gets updated and the computed biasing
forces are added to the backend net forces.

Finally, the information collected by the sampling method is returned and can be used for calculating the free energy as function
of the selected \acp{CV}. Unlike \ac{SSAGES}, \PySAGES offers a user-friendly \texttt{analyze} interface that simplifies the process of performing post simulation analysis, including the automatic calculation of free energies based the chosen sampling method. This feature can greatly reduce the time and effort required to gain valuable insights from simulations.

\PPySAGES offers an easy way to leverage different parallelism
frameworks including \ac{MPI} with the same uniform 
fronted available to run enhanced sampling simulations.
This is achieved via Python's \texttt{concurrent.futures}
interface. In particular, for \ac{MPI} parallelism, the
user only needs to pass an additional \texttt{MPIPoolExecutor}
(from \texttt{mpi4py}) to \PySAGES' \texttt{run} method.
If the user selects a method such as \texttt{UmbrellaSampling},
the workload for each image will be distributed across
available \ac{MPI} nodes. On the other hand, for most of the
sampling methods, the parallelization interface allows the
user to run multiple replicas of the same system to enable,
for instance, analysis of the uncertainties associated to computing the free energy of a given system.

To ensure the reproducibility and correctness of our implementation and to follow software engineering best practices, we have implemented a
comprehensive unit tests suite, and leverage GitHub's
continuous integration services. In addition, we use
\url{trunk.io}\;\cite{trunk} to adhere to quality
standards as well as to ease the collaboration of developers.

\subsection{Enhanced Sampling Methods}

While we assume the reader has some basic understanding of enhanced sampling methods, here we provide an overview of these techniques. We direct readers interested in learning more about the fundamentals of enhanced sampling to a number of excellent recent review articles\;%
\cite{sidky2020machine,jackson2019recent,tiwary2016review,wang2019advances,mitsutake2013enhanced,miao2016unconstrained,yang2019enhanced,bussi2014,limongelli2020,catalysis}. In addition, we discuss the general structure of how enhanced sampling methods are implemented within \PySAGES, and also present a summary of the various methods already available in the library.

Enhanced sampling methods are a class of simulation techniques that manipulate
regular \ac{MD} simulations in order to more effectively sample the configuration space.
In \ac{MD} a \acl{CV}, $\xi$, is typically a function of the positions of all particles,
$\hat\xi(\{r_i\})$.

For a given statistical ensemble (such as the canonical, \caps{NVT}), the corresponding 
free energy can be written as $A=-k_\text{B}T\ln(Z)$, where $A$ is 
the Helmholtz free energy and $Z$ is the canonical partition function.
To make explicit the dependency of the free energy on $\xi$, let us write down the partition function:
\begin{align}
    Z(\xi) \propto \int \text{d}^N r_i\;\delta(\hat{\xi}(\{r_i\})-\xi)\;e^{-U(\{r_i\})/k_\text{B}T}\label{eq:partition-function}
\end{align}

Normalizing this partition function gives us 
the probability of occurrence, $p(\xi) = Z(\xi)/(\int\text{d}\xi\; Z(\xi))$,
for configurations in the \ac{CV} subspace.
Substituting this probability into the expression for the free energy, we get:
\begin{align}
    A(\xi) = -k_\text{B}T \ln(p(\xi)) + C\label{eq:free-energy}
\end{align}
where $C$ is a constant.

If we take the derivative of the free energy with respect to
$\xi$ we get
\begin{align}
    \frac{dA(\xi)}{d\xi} =
    \frac{
        \int \text{d}^N r_i\; \frac{dU}{d\xi}\;
        \delta(\hat{\xi}(\{r_i\})-\xi)\;
        e^{-U(\{r_i\})/k_\mathrm{B}T}
    }{
        \int \text{d}^N r_i\; \delta(\hat{\xi}(\{r_i\})-\xi)\;
        e^{-U(\{r_i\})/k_\mathrm{B}T}
    } =
    \bigg\langle\frac{dU}{d\xi}\bigg\rangle_{\!\!\xi},%
    \label{eq:fe-derivative}
\end{align}
where $\langle\ldots\rangle_\xi$ denotes the conditional average.

The goal of \ac{CV}-based enhanced sampling methods is to accurately determine either $p(\xi)$ or $dA(\xi)/d\xi$\emdash from which $A(\xi)$ can be recovered\emdash in a computationally tractable manner.

In \PySAGES, the implementation of sampling methods follows the \JAX functional style programming model. New methods are implemented as subclasses of the \texttt{SamplingMethod} class, and are required to define a \texttt{build} method. This method returns two methods, \texttt{initialize} and \texttt{update}, used as part of the process of biasing the simulation. For readers familiar with \JAXMD, these could be thought of as analogues to the higher level functions returned by \JAXMD's \texttt{simulate} integration methods. The \texttt{initialize} method allocates all the necessary helper objects and stores them in a \texttt{State} data structure, while the \texttt{update} method uses the information from the simulation at any given time to update the \texttt{State}.

While \PySAGES allows new methods to be written seamlessly as part of Python scripts used to set up molecular dynamics simulations, it also provides out-of-the-box implementations of several of the most important known sampling methods. We list and briefly detail them next.

\subsubsection{Harmonic Biasing}\label{sec:harmonic-bias}

One simple way to sample a specific region of the phase space is to bias the simulation around a point $\xi_0$ with harmonic bias.
This adds a quadratic potential energy term to the Hamiltonian that increases the potential energy as a system moves away from the target point: $\mathcal{H}_b = \mathcal{H} + k/2 (\xi - \xi_0)^2$,
where $k>0$ is the spring constant.
The unbiased probability distribution $p(\xi)$ can be recovered by dividing the biased distribution by the known weight of the bias $p(\xi) = p_b(\xi)/e^{-k/2(\xi-\xi_0)^2/k_\text{B}T}$.

The disadvantage of this approach is that it can only be used to explore the free energy landscape near a well-know point in phase space.
This may not be sufficient for many systems, where the free energy landscape is complex.

\subsubsection{Umbrella Sampling}\label{sec:umbrella-integration}

Umbrella sampling is a technique that traditionally builds on harmonic biasing by combining multiple harmonically-biased simulations.
It is a well-known method for exploring a known path in phase space to obtain a free energy profile along that path\;\cite{kastner2009umbrella,kastner2011umbrella}.
Typically, a path between to point of interest is described by $N$ points in phase space, $\xi_i$.
At each of these points, a harmonically biased simulation is performed, and the resulting occurrence histograms are combined to obtain a single free energy profile.

In \PySAGES, we implement umbrella integration for multi-dimensional \acp{CV}.
This method approximates the forces acting on the biasing points and integrates these forces to find the free energy profile $A(\xi)$, and allows to explore complex high-dimensional free energy landscapes.

\subsubsection{Improved String Method}\label{sec:string}

When only the endpoints are known, but not the path itself, the improved (spline-based) string method can be used to find the \ac{MFEP} between these two endpoints\;\cite{weinan2007simplified}.
The spline-based string method improves upon the original string method by interpolating the \ac{MFEP} using cubic-splines.
In this method, the intermediate points of the path are moved according to the recorded mean forces acting on them, but only in the direction perpendicular to the contour of the path. This ensures that distances between the points along the path remain constant.

This method has been widely used and has been shown to be an effective way to find the \ac{MFEP} between two points in the phase space\;\cite{weinan2007simplified}.

\subsubsection{Adaptive Biasing Force sampling}\label{sec:abf}

The \ac{ABF} sampling method is a technique used to map complex free-energy landscapes. It can be applied without prior knowledge of the potential energy of the system, as it generates on-the-fly estimates of the derivative of the free energy at each point along the integration pathway. \Ac{ABF} works by introducing an additional force to the system that biases the motion of the atoms, with the strength and direction of the bias continuously updated during the simulation. In the long-time limit, this yields a Hamiltonian with no average force acting along the transition coordinate of interest, resulting in a flat free-energy surface and allowing the system to display accelerated dynamics, thus providing reliable free-energy estimates\;\cite{comer2015adaptive,darve2008adaptive}.
Similarly to \ac{SSAGES}, \PySAGES implementation of \ac{ABF} is based on the algorithm described in\;\cite{darve2008adaptive}.

\subsubsection{Metadynamics}

Metadynamics is another popular approach for enhancing sampling of complex systems. In metadynamics\;\cite{laio2002escaping}, a bias potential is applied along one or more \acp{CV} in the form of Gaussian functions.
The height and width ($\sigma$) of these Gaussians are controlled by the user. 
The Gaussian bias potentials are cumulatively deposited at user-defined intervals during the simulation.
In standard metadynamics, the height of the Gaussian bias potentials is fixed.

In contrast, for \ac{WTMD}\;\cite{barducci2008well} simulations, the height of the Gaussian bias potentials is adjusted at each timestep using a preset temperature based bias factor.
This scaling of Gaussian heights in \ac{WTMD} leads to faster convergence compared to standard metadynamics, as it restricts the range of free energy explored to a range defined by the bias factor.

In \PySAGES, we have implemented both standard metadynamics and \ac{WTMD}. The well-tempered variant is activated when a user sets a value for the bias factor. 
To improve the computational performance, we have added optional support for 
storing the bias potentials in both on a pre-defined grid. This allows users to trade-off accuracy for faster simulations, depending on their needs.

\subsubsection{Forward Flux Sampling}

\Ac{FFS} belongs to a different family of enhanced sampling methods than the ones described above. In the previously described methods, the free energy change from a region in the phase space ($A$) to the region of interest ($B$) is calculated by applying a bias to the system. In \ac{FFS} no bias is added and instead an efficient selection of trajectories that crosses the phase space from $A$ to $B$ is performed. Since no bias is used, the intrinsic dynamics of the system is conserved and therefore kinetic and microscopic information of the transition path can be studied\;\cite{ffs3}. In \PySAGES we have implemented the direct version of \ac{FFS}\;\cite{ffs1,ffs2}.

\subsubsection{Artificial neural networks sampling}

\Ac{ANN}\;\cite{2018HSidky-JCP-ANN} employs regularized neural
networks to directly approximate the free energy from the
histogram of visits to each region of the \ac{CV} space, and
generates a biasing force that avoids ringing and boundary
artifacts\;\cite{2018HSidky-JCP-ANN}, which are commonly
observed in methods such as metadynamics or basis functions
sampling\;\cite{whitmer2014basis}. This approach is effective
at quickly adapting to diverse free energy landscapes by
interpolating undersampled regions and extrapolating bias
into new, unexplored areas.

The implementation on \PySAGES offers more flexible approaches to network regularization than \ac{SSAGES}, which uses Bayesian regularization.

\subsubsection{Force-biasing using neural networks}

\Ac{FUNN}\;\cite{2018AZGuo-JCP-FUNN} is based upon the same
idea as \ac{ANN}, that is, relying on artificial neural
networks to provide continuous functions to bias a simulation,
but instead of using the histogram to visits to \ac{CV} space
it updates its network parameters by training on the \ac{ABF}
estimates for the mean forces as the simulation advances.
This method shares all of the features of \ac{ABF}, but the
smooth approximation of the generalized mean force it produces
enables much faster convergence to the free energy of a system
compared to \ac{ABF}.

\subsubsection{Combined Force Frequency sampling}

The \ac{CFF} method\;\cite{2020ESevgen-JCTC-CFF} combines the speed of generalized-force based techniques such as \ac{ABF} or \ac{FUNN} with the advantages of frequency-based methods like metadynamics or \ac{ANN}. Notable improvements over earlier force-based methods include eliminating the need for hyperparameters to dampen early-time estimates, automating the integration of forces to generate the free energy, and providing an explicit expression for the free energy at all times, enabling the use of replica exchange or reweighing.

In principle, by using sparse storage of histograms, it should
be possible to scale the method to higher dimensions without encountering memory limitations, such optimization is however
not yet implemented in \PySAGES.

\subsubsection{Spectral Adaptive Biasing Force}\label{sec:spectral}

Spectral~\ac{ABF}\;\cite{2022PFZubietaRico-arxiv} is a method
that follows the same principle as neural-network-based
sampling methods, in that it builds a continuous approximation
to the free energy. However, in contrast to methods like
\ac{FUNN} it does so by fitting exponentially convergent
basis functions expansions, and could be thought as a
generalization of the Basis Functions Sampling Method.
In contrast to the latter, and similar to \ac{CFF},
it allows for the recovery of an explicit expression for
the free energy of a system. It is an extremely fast method
in terms of both runtime and convergence.

\subsection{Collective variables}

As previously mentioned, enhanced sampling calculations
commonly involve the selection of a \ac{CV}. An appropriate
\ac{CV} for a given system could simply be the distance
between the centers of mass of two groups of atoms, but could
be a complex specialized quantity.

Below, we list a set of \acp{CV} predefined in \PySAGES,
sorted by the number of groups of atom coordinates necessary
for their use:
\begin{enumerate}\raggedright
    \item \texttt{TwoPointCV}. This subclass is for \acp{CV}
    that need two groups for their definition. This includes
    \texttt{Distance} and \texttt{Displacement} (vector).

    \item \texttt{ThreePointCV}. Subclass of \acp{CV} with
    three groups of atoms, such as \texttt{Angle}.

    \item \texttt{FourPointCV}. Subclass of \acp{CV} with four
    groups of atoms, such as \texttt{DihedralAngle}.

    \item \texttt{AxisCV}. Subclass of \acp{CV} that are
    projected on a determinate axis. This includes
    \texttt{Component} and \texttt{PrincipalMoment}.

    \item \texttt{CollectiveVariable} General base class for
    all \acp{CV}. In \PySAGES, \acp{CV} that directly
    derive from this class, and do not belong to the previous
    groups, include: \texttt{RingPhaseAngle},
    \texttt{RingAmplitude}, \texttt{RadiusofGyration},
    \texttt{Asphericity}, \texttt{Acylindricity},
    \texttt{ShapeAnisotropy},
    \texttt{RingPuckeringCoordinates}\;\cite{puckering}
    (vector).
\end{enumerate}

\begin{figure}[htp]
  \centering
  \includegraphics[width=\textwidth]{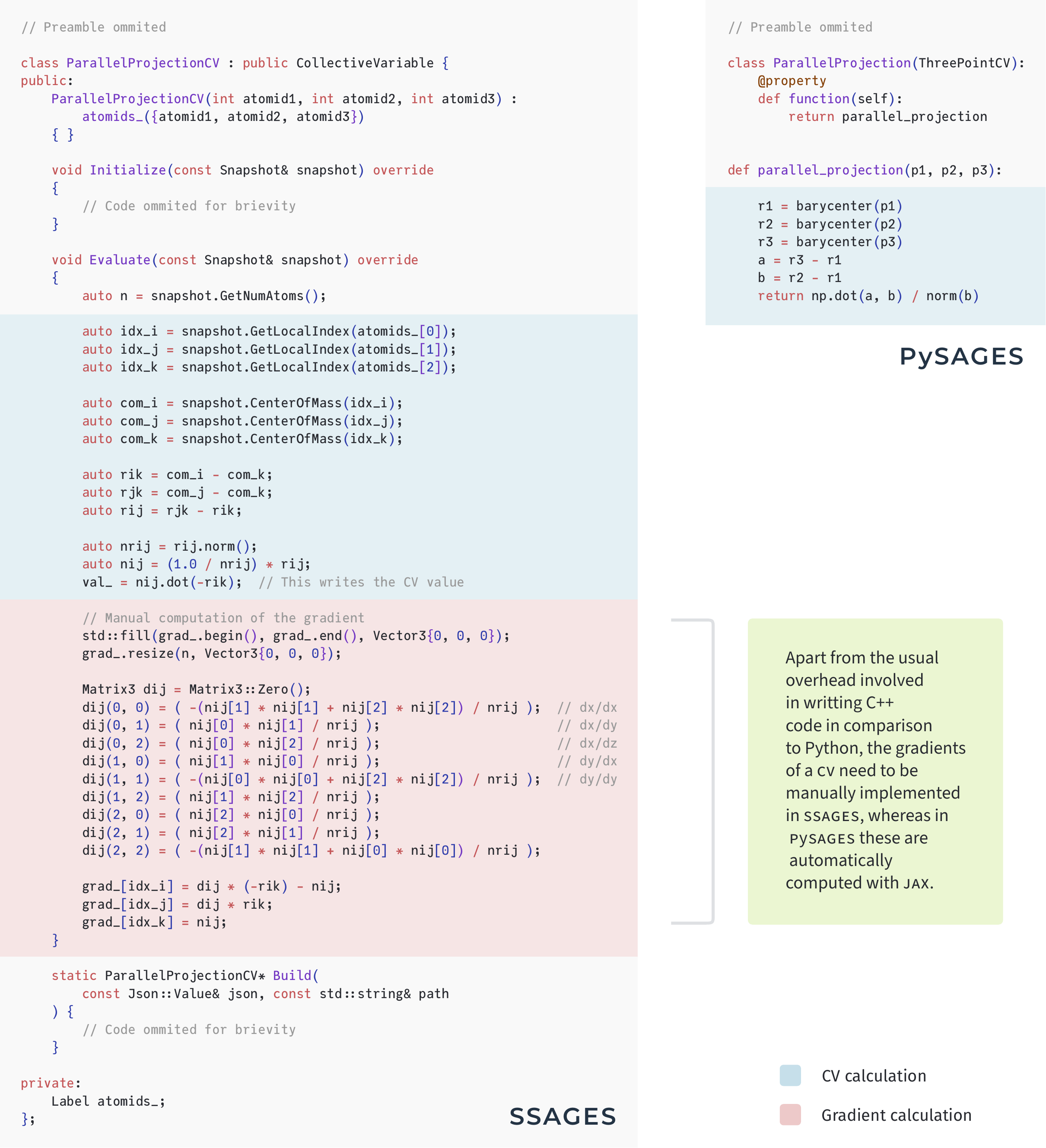}
  \caption{
    Example of how to write a \ac{CV} in \PySAGES. On the left is the same \acp{CV} written in \ac{SSAGES} and on the right the \PySAGES version. 
    In general, the only requirement is for the user to write the \ac{CV} as a differentiable function in \JAX.
  }
  \label{fig:cv-comparison}
\end{figure}

In \PySAGES we provide users with a simple framework for
defining \acp{CV}, which are automatically differentiated with
\JAX. To illustrate this, we compare how to write the
calculation of a \ac{CV} that measures the projection of the
vector between two groups of atoms over the axis that passes
by other two groups, in both \ac{SSAGES} and \PySAGES
(see \autoref{fig:cv-comparison}). In \PySAGES the gradient
calculation is done automatically whereas in \ac{SSAGES} it
has to be coded explicitly.

Data-driven and differentiable \acp{CV} discovered using artificial neural networks (e.g. autoencoders)\;\cite{sidky2020machine, chen2019nonlinear, chen2018molecular, ribeiro2018reweighted, wehmeyer2018time} with arbitrary featurizations of atoms can in principle be implemented in \PySAGES based on the above general abstract classes of \acp{CV}.

The following second example shows the power of differential programming for \ac{CV} declaration in \PySAGES.

\subsubsection{Case study: A \acl{CV} for interfaces}%
\label{sec:cvs-interface}
When the two immiscible liquids are in contact with each other, the density of one liquid experiences a gradual change. This transition region is the liquid-liquid interface and its position has high importance in many studies (see \autoref{sec:LC}). However, the location of such interface is not a trivial task since it generally fluctuates as the simulation progresses. As a representative \ac{CV} for the interface, we can utilize the position of the point where the gradient of the density is maximized. More formally, let $\rho(x)$ denote the density of a liquid of interest at a coordinate $x$ on the perpendicular axis. We would like to find the location of the interface:
\begin{equation}
I = \arg\max_{x} |\rho'(x)|
\end{equation}

However, the density function $\rho(x)$ is not directly measurable in a molecular simulation, as the coordinates of atoms are discrete. To obtain an approximation of $\rho(x)$, we divide the coordinates into multiple bins, each with a width of $\delta$, and create a histogram $p(x)$ that records the number of atoms falling into the bin around position $x$. In other words,
\begin{equation}
p(x) = \sum_{i=1\ldots n}\big[\,|x_i - x| < \delta / 2\,\big]
\end{equation}
in which $x_i$ denotes the coordinate of atom $i$. As written above, $p(x)$ is non-differentiable. Therefore, as in other works\;\cite{cvdiff}, we utilize the kernel density trick with a Gaussian kernel to modify $p(x)$. The modified $\tilde{p}(x)$, is defined as:
\begin{equation}
\tilde{p}(x) = \sum_{i=1\ldots n}\exp{\left(-\frac{(x_i-x)^2}{2\sigma^2}\right)}
\end{equation}
in which $\sigma$ is a hyperparameter that decides the width of the Gaussian kernel. Then, the gradient of the density can be approximated as: 
\begin{equation}
\tilde{p}\,'(x) = \frac{\tilde{p}(x+\delta/2)-\tilde{p}(x-\delta/2)}{\delta}    
\end{equation}
and we calculate the location of the interface as $I = \arg\max_{x}|\tilde{p}\,'(x)|$.
The $\arg\max$ operator is also non-differentiable. As a result, we replace it with a softmax function that transforms the raw input into a probability. Denote the $m$ bins as $j = 1\ldots m$, and finally we calculate the location of the interface as:
\begin{equation}
I = \frac{\sum_{j}x_j\,\exp|\tilde{p}\,'(x_j)|}{\sum_{j}\exp|\tilde{p}\,'(x_j)|}    
\end{equation}

As demonstrated in the code snippet for this \ac{CV}, provided
in \autoref{appendix:interface}, \PySAGES allows for the concise
and straightforward implementation of complex \acp{CV} such as
this one.

\section{Results and Discussion}\label{sec:results}

To evaluate a software package like \PySAGES, we must
consider at least two factors: physical correctness and
computational performance.

First, to assess the correctness of the enhanced sampling
methods implemented in \PySAGES, we present in
\ref{appendix:alanine} the free-energy landscape for the dihedral
angles $\phi$ and $\psi$ of \ac{ADP}. This example
is commonly used to benchmark new enhanced sampling
algorithms. Similarly, we also show in \ref{appendix:butane} the
free-energy as a function of the dihedral angle of butane.
Our results show that \PySAGES reproduces the expected
free-energy landscapes using different methods and backends.
In \autoref{sec:example-systems}, we further investigate the
applicability and correctness of \PySAGES beyond these simple
model systems.

Second, we demonstrate the performance of \PySAGES on \acp{GPU}
with two different backends in \autoref{sec:performance}.
In particular, we compare the performance of enhanced sampling
simulations to the performance of pure \ac{MD} simulations,
as well as other enhanced sampling implementations.

\subsection{Example applications of enhanced sampling with \texorpdfstring{\upshape{\PySAGES}}{PySAGES}}\label{sec:example-systems}

To demonstrate the versatility and effectiveness of \PySAGES
in different contexts, we present several examples of how
enhanced sampling methods can be used to gain valuable
insights in various fields including biology, drug design,
materials engineering, polymer physics, and ab-initio
simulations. These examples showcase how \PySAGES can be
used in diverse research areas and the utility of different
enhanced sampling methods and backends.

Overall, these examples confirm that the enhanced sampling methods implemented in \PySAGES work as intended and provide results consistent with existing literature.

\subsubsection{Structural Stability of Protein--Ligand Complexes for Drug Discovery}
High-throughput docking techniques are a widely-used computational technique in drug lead discovery. 
However, these techniques are limited by the lack of information about protein conformations 
and the stability of ligands in the docked region\;\cite{sethi2019molecular}.
To address this issue, the Dynamical Undocking (\DuCK) method was developed to evaluate the stability of the ligand binding by calculating the work required to break the most important native contact (hydrogen bond interactions)  in the protein-ligand complex\;\cite{duck}.
This method has been shown to be complementary and orthogonal to classical docking, making both techniques work parallel in drug discovering\;\cite{ducktest1,ducktest2}. 
However, \DuCK can be slow to converge when combined with
traditional enhanced sampling techniques\;\cite{duck},
making it unsuitable for high-throughput drug discovery protocols.

\begin{figure}[htbp]
    \centering
    \includegraphics[width=0.75\textwidth]{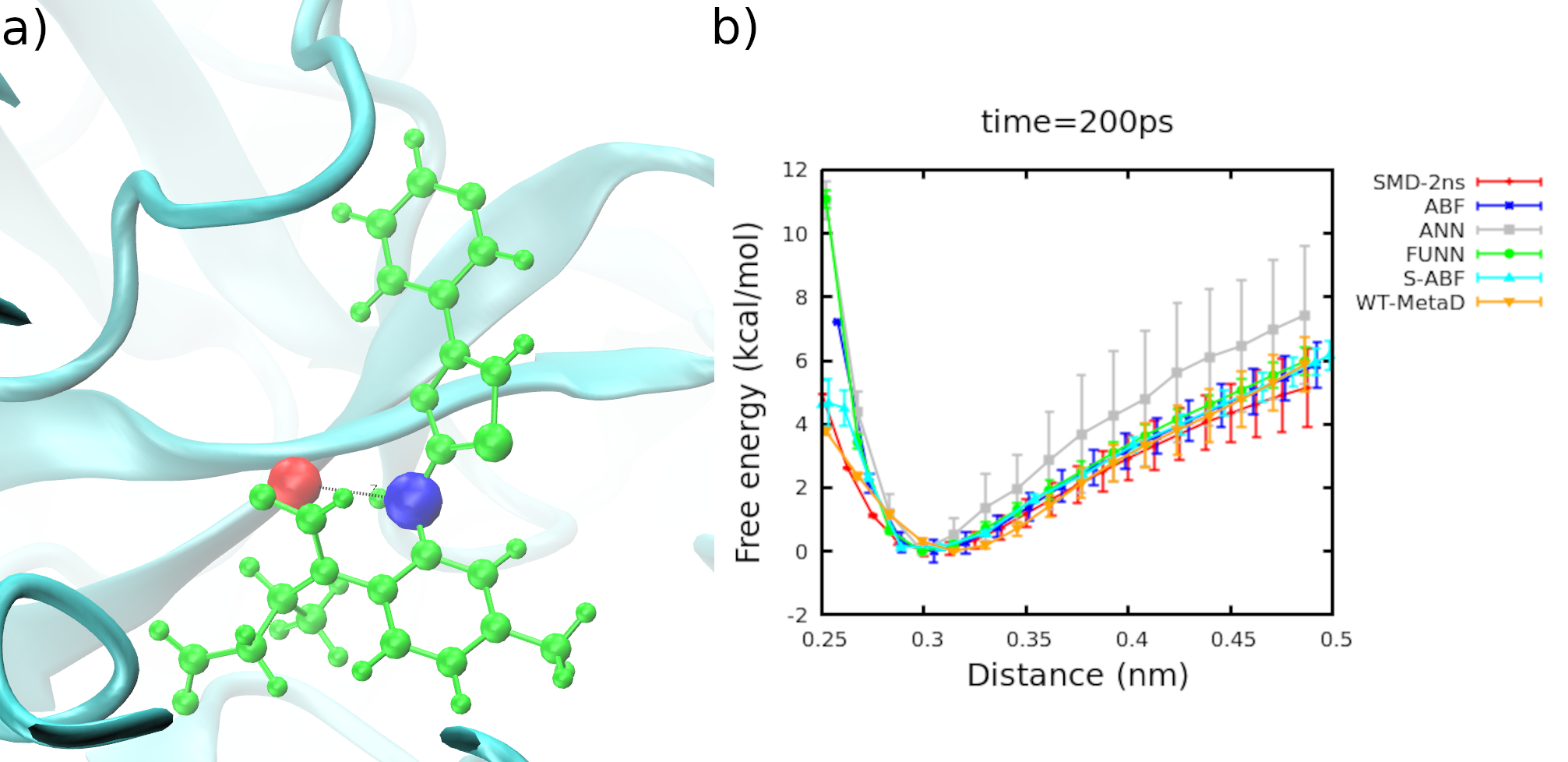}
    \caption{Dynamical Undocking (\DuCK) method in detail.
    For a proposed binding mode obtained from classical docking, a short run using \ac{MD} simulations is carried out and the most stable receptor-ligand native contact is selected from that run.
    In this case, it is the hydrogen bond between the red and blue atoms highlighted in panel a). 
    b) Comparison between different methods in \PySAGES for \DuCK calculations
    averaged over 5 different replicas for each method. 
    The reference, a Steered \ac{MD} simulations simulations of 2 ns is in red. 
    In comparison, different methods in \PySAGES are used considering simulation period 10 times shorter: 
    only \ac{ANN}\;\cite{2018HSidky-JCP-ANN} provides inferior performance 
    against the reference; Spectral~\ac{ABF}\;\cite{2022PFZubietaRico-arxiv} 
    or \ac{FUNN}\;\cite{2018AZGuo-JCP-FUNN} give the best performance.}
    \label{fig:biological2}
\end{figure}

Here, we demonstrate how \PySAGES with \OpenMM can be used efficiently in drug discovery applications, where the user-friendly interface, native parallel capabilities, and new enhanced sampling methods with fast convergence are synergistically combined to accelerate the virtual screening of ligand databases.
In this example, we study the main protease (Mpro) of Sars-CoV-2 virus 
(\caps{PDB: 7JU7}\;\cite{duck0}), where the ligands were removed and the monomer A was selected as the docking receptor.
A ligand with \ac{SMILES} string \texttt{CCCCOCC(=O)c1ccc(C)cc1N[C@H]1N[C@@H](c2cccnc2)CS1} was docked using RDock\;\cite{rdock}.
The best scoring pose was used to initialize 
the system, which was simulated using the
\textls[60]{ff}\kern0em\caps{14SB}\;\cite{ff14sb},
\caps{TIP3P}\;\cite{1983WLJorgensen-JChemPhys},
and \caps{GAFF}\;\cite{2004JWang-JComputChem} force fields. 
A 10 ns equilibration procedure was carried out to find the most stable hydrogen bond between the ligand and the protein.
The last frame of this equilibration was then used to initialize 
the enhanced sampling calculations in \PySAGES with \ac{ABF}, metadynamics, \ac{FUNN}, \ac{ANN}, and Spectral~\ac{ABF}.
These methods were compared against the same system simulated using Amber20\;\cite{amber20} with Steered Molecular Dynamics (see \autoref{fig:biological2}b). 
Our results suggest that we can reduce the simulation time by an order of magnitude using new enhanced sampling methods like Spectral~\ac{ABF} or \ac{FUNN}.
This can greatly accelerate the drug discovery process and help identify potential drug leads more quickly. 

\subsubsection{Fission of a Diblock Copolymer Spherical Domain}\label{sec:polymer}

\begin{figure}
    \centering
    \includegraphics[width=\textwidth]{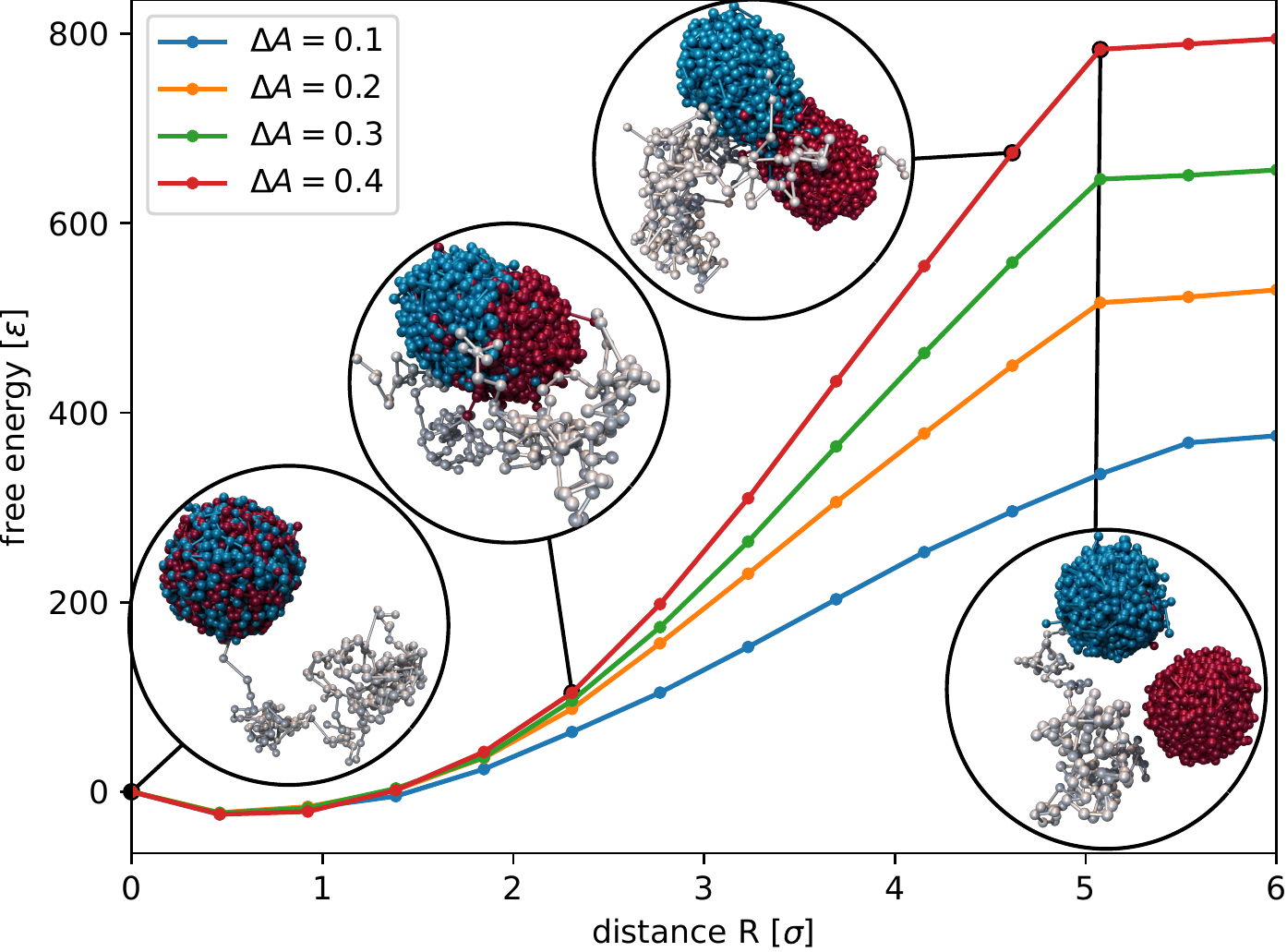}
    \caption{Free energy landscape of the fission of a spherical diblock-copolymer domain. 
    The chain ends forming the spherical domain are split into two groups (blue) and (red), 
    the other chain ends not visible for clarity except for a single chain (grey).
    Initially, a single spherical domain is formed, but as we constraint the center of mass 
    between the blue and red groups further, the domain first elongates 
    and then separates completely. During this separation, the free energy continuously increases 
    and the increase is steeper for high repulsion between unlike type $\Delta A$. 
    As soon as the domain is separated, the free energy plateaus.}
    \label{fig:polymer}
\end{figure}

We now investigate the fission of a single spherical domain of a diblock copolymer using a coarse-grained model.
We use a soft, coarse-grained \ac{DPD} model published in previous studies\;\cite{schneider2018transitions,schneider2019rheology,schneider2020symmetric}.
The model consists of $n=200$ chains with $N=256$ beads each, representing a liquid polymer melt.
The first $N_A=16$ beads in each chain are type A, while the remaining $N_B=240$ are type B.

A standard \ac{DPD} potential is used to enforce incompressibility with a repulsion parameter of
$A_{ii}=5k_BT/\sigma^2$.
However, a higher interaction of $A_{AB} = A_{ii} + \Delta A k_BT/\sigma^2$, 
with $\Delta A \in [0.1,0.4]$ is applied between unlike particles to create a repulsion that leads to a microphase separation. A Flory-Huggins parameter $\Delta A \propto \chi N > 0$ can characterize this phase separation.
The interaction range of this non-bonded potential is $1\sigma$, as well as 
the range of the \ac{DPD} thermostat that keeps the temperature at $T=1k_BT=1\epsilon$.

In addition, a harmonic spring force with zero resting length is used to connect the beads to polymer chains with a spring constant of $k=16/3k_B/\sigma^2$, resulting in an average bond length of $b_0=0.75\sigma$.
The equilibrium phase for this polymer melt is a \ac{BCC} phase of spherical A droplets inside a B melt.\;\cite{matsen2001standard}
However, we confine the polymer to a tight cubic simulation box of length $L_0 = 10\sigma$, which results in a single A spherical domain in the B matrix.
We integrate the simulation with a time step of $\Delta t = 10^{-3}\tau$ and each simulation is equilibrated for $t=1000\tau$, followed by a production run of $t=1000\tau$ as well.
A discussion of the \ac{GPU} performance of this system with and without \PySAGES can be found in \autoref{sec:gpu-performance}.

After defining the diblock copolymer system, the next step is to define a \ac{CV} within the system. 
In this case, we are interested in the fission of the single spherical A domain into two equally sized smaller A domains. 
To achieve this, we divide the polymer chains into two groups:
the first $n=100$ chains are going to form the first small domain (blue in \autoref{fig:polymer})
and the second $n=100$ chains form the second spherical domain (red in \autoref{fig:polymer}).
To define and enforce the separation of the two groups, we define our \ac{CV} as the distance, $R$, between the center of mass of the blue A-tails and the center of mass of the red A-tails.
Initially, without biasing, the two groups form a single spherical domain 
and blue and red polymer tails are well mixed, as shown at small $R<1\sigma$ in \autoref{fig:polymer}.

To study the separation of the spherical domain, we use harmonic biasing 
(see \autoref{sec:harmonic-bias}) to enforce a separation distance $R_0$ between the two groups.
The high density in the system $\sqrt{\text{\textsmaller{$\bar{\mathcal{N}}$}}} = \frac{\rho_0}{N} R_{e0}^3 \approx 344$, leads to low fluctuations and suppression of unfavorable conformations.
Therefore, we use a high spring force constant of $k_{CV} = 1500\epsilon/\sigma^2$ to facilitate the separation.

We investigate a separation of $R\in[0,6]\sigma$ with 14 replicas and use umbrella integration
(see \autoref{sec:umbrella-integration}) to determine the free energy profile, as shown in \autoref{fig:polymer}.
As we increase the external separation distance $R_0$, we observe how the single domain splits into two.
At a low separation distance $R<2\sigma$, the single domain is mostly undeformed, but the two groups separate inside the single spherical domain.
Increasing the separation distance further goes beyond the dimensions of the spherical domain, leading to the deformation of the domain into an elongated rod-like shape.
The two groups still maintain a connection to minimize the AB interface.

At a separation between $4\sigma$ and $5\sigma$ the deformation becomes so strong, that the penalty of forming another AB interface between the two groups, and hence forming two spherical domains, is lower than the entropic penalty of the domain deformation and elongated AB interface of the droplet.
After the separation, the free energy landscape remains indifferent to the separation, since there is no interaction between the two domains left.

The free energy profile of separation is controlled by the repulsion of unlike types 
$\chi N \propto \Delta A$. The stronger the repulsion, the more energy is necessary 
to enlarge the AB surface area for the fission.
For the strongest interaction $\Delta A = 0.4\epsilon$, the total free energy barrier reaches 
about $800\epsilon$, while for the lowest $\Delta A=0.1\epsilon$ it remains below $400\epsilon$.
Both barriers are orders of magnitude larger than thermal fluctuations $1k_BT=1\epsilon$, so a spontaneous separation is not expected and the fission can only be studied via enhanced sampling.

It is interesting to note that at the lowest separation distance $R_0=0$ it is not the lowest free energy state.
Enforcing perfect mixing is not favorable, as the two groups naturally want to separate slightly optimizing the entropy of the chain end-tails.

\subsubsection{Liquid Crystal Anchoring in Aqueous Interfaces}\label{sec:LC}
\begin{figure}[htbp]
    \centering
    \includegraphics[width=0.5\textwidth]{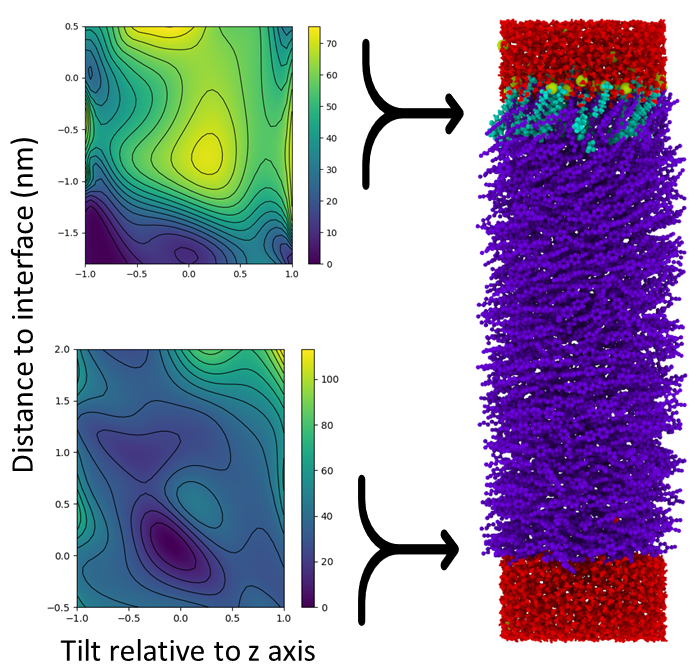}
    \caption{Free energy surface of \ac{5CB} in a hybrid anchoring slab with \ac{SDS} and water. Right: Snapshot of the system with water molecules in red, \ac{5CB} in purple, \ac{SDS} in green and sodium ions in yellow. Top Left: \ac{FES} of \ac{5CB} molecule near the water--\ac{SDS} interface. Bottom Left: \ac{FES} of \ac{5CB} near a pure water interface. Both \ac{FES} were obtained with \PySAGES and \OpenMM using the \ac{FUNN} method.}
    \label{fig:LC-material}
\end{figure}

\Acp{LC}, materials that flow like liquids but have anisotropic properties as crystals, have been used lately as prototypes for molecular sensors at interfaces given the high sensitivity in their anchoring behavior relative to small concentration of molecules at aqueous interfaces\;\cite{LC-rev}. The presence of molecules at the interface changes drastically the free energy surface of \ac{LC} molecules relative to their orientation and distance to such interface. In this example, we are revisiting some canonical interfaces for \ac{LC}; \ac{5CB} at the interface of pure water and \ac{SDS}. For \ac{5CB} and water, previous work has focused on obtaining the free energy surface of a \ac{5CB} at the water interface\;\cite{hadi}. In our case, hybrid anchoring conditions have been imposed on a 16~nm slab of 1000 \ac{5CB} molecules in the nematic phase (300~K) interacting with a 3~nm slab of water with 62 molecules of \ac{SDS} at one of the interfaces. The force fields used are: \emph{united atom} for \ac{5CB}\;\cite{5cb}, \caps{TIP3P}\;\cite{1983WLJorgensen-JChemPhys} for water, \caps{GAFF}\;\cite{2004JWang-JComputChem} and Lipid 17 for \ac{SDS}. The \acp{CV} chosen to study this system are the distance of the center of mass of one molecule of \ac{5CB} at each one of the interfaces (see~\ref{appendix:interface}), and the tilt orientation of the same molecule with respect to the z axis of the box. The free energy surfaces for the pure water and with \ac{SDS} at the interface are both displayed in \autoref{fig:LC-material}. We can observe that the free energy surface of pure water shows a minimum corresponding to a parallel orientation to the surface with a similar shape that one calculated in\;\cite{hadi}. On the contrary, the presence of \ac{SDS} transforms the minimum to a maximum in the same relative position and orientation to the interface (\autoref{fig:LC-material} top left), moving now the minima to a perpendicular orientation of \ac{5CB} to the interface, in agreement to the experimental observation of change from planar to homeotropic anchoring in the presence of \ac{SDS} in water.   

\subsubsection{Ab Initio Enhanced Sampling Simulations}
In the field of \textit{ab initio} simulations of heterogeneous catalysis, capturing the dynamic and entropic effects is crucial for an accurate description of the phenomena\;\cite{catalysis}.
Classical force fields are inadequate for capturing the essential bond breaking events involved in catalysis, so \ac{MD} simulations based on first-principles calculations are necessary.
Given that reactive events are often limited by large free energy barriers, enhanced sampling techniques are a crucial part of these simulations.
Coupling \PySAGES to \ac{ASE}, provides access to a wide range of first-principle calculators.

\begin{figure}[htbp]
    \centering
    \includegraphics[width=0.8\textwidth]{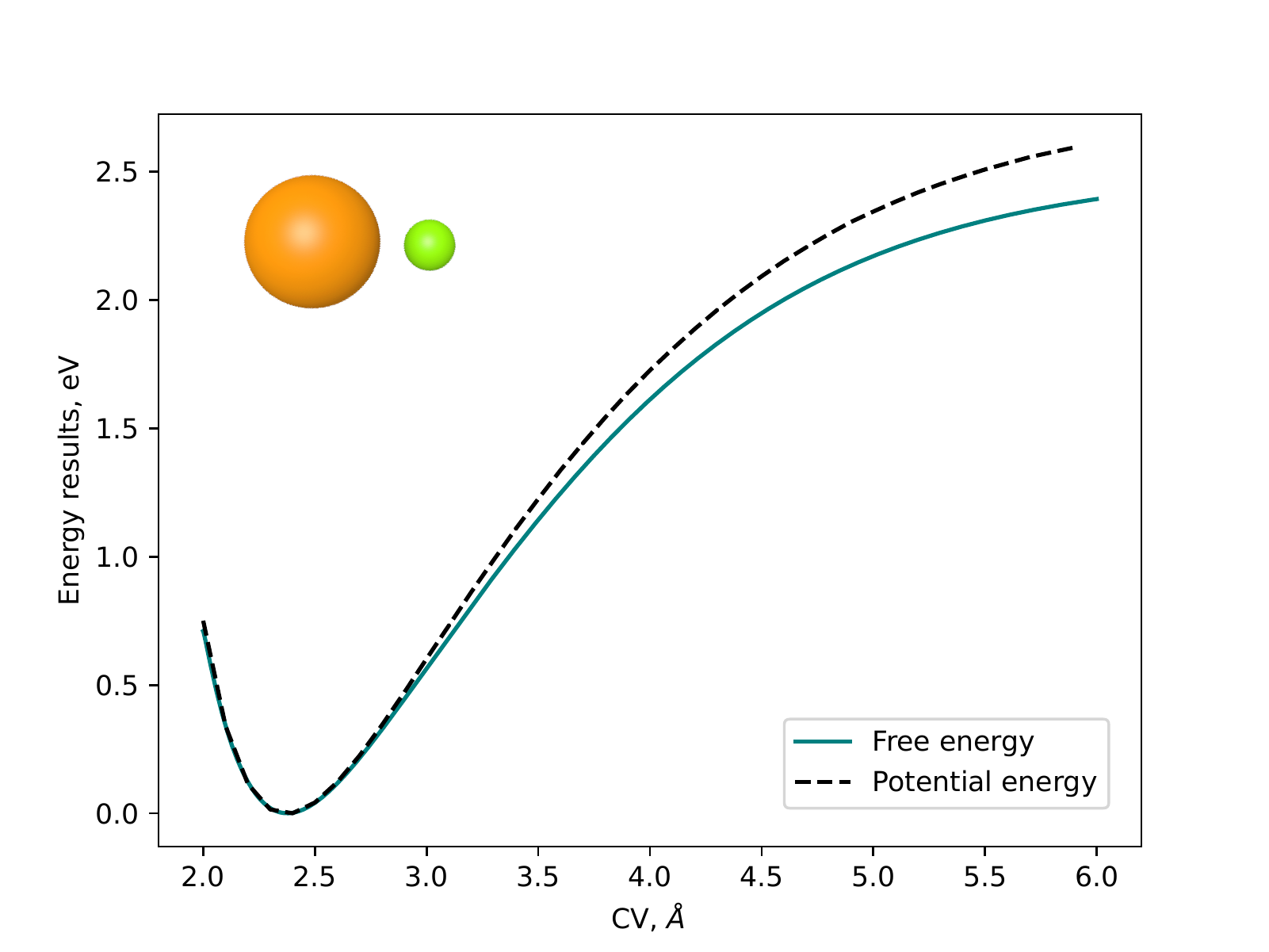}
    \caption{Free energy ($T = 300$~K) and potential energy calculations of Na--Cl distance with \ac{ASE}~+~\caps{VASP} using Spectral~\ac{ABF} in \PySAGES.}
    \label{fig:abinitio1}
\end{figure}

As an example, we have used \caps{VASP} as a calculator for a simple 
\textit{ab initio} enhanced sampling simulation.
The \ac{CV} is the separation distance between a sodium and chlorine atom using the \caps{PBE} functional\;\cite{PBE},  and Spectral~\acs{ABF} as the enhanced sampling method (see \autoref{sec:spectral}).
The results are shown in \autoref{fig:abinitio1}, where the minimum in the free energy profile along the Na--Cl distance corresponds to the equilibrium distance between Na and Cl atoms in vacuum.

\subsubsection{Enhanced Sampling with Machine Learning Force Fields}
Deep \ac{NN} force fields can retain the accuracy of \textit{ab initio} \ac{MD} while allowing for computational costs similar to those of classical \ac{MD}. 
Through \ac{ASE} it is possible to access \ac{NN} potentials such as \DeepMD\;\cite{deepmd}, and the \ac{GAP}. 
Additionally, \JAXMD allows to leverage more general \ac{NN} potentials 
that can be used in enhanced sampling calculations. 
Coupling of \PySAGES with \ac{ASE} or \JAXMD can be used in active learning of \ac{NN} force fields by efficiently sampling rare events using any of the enhanced sampling methods provided by \PySAGES as described in Ref.\;\cite{deepmeta} where parallel tempering metadynamics 
was used to generate accurate \ac{NN} force field in urea decomposition in water. 

\begin{figure}[htbp]
    \centering
    \includegraphics[width=0.9\textwidth]{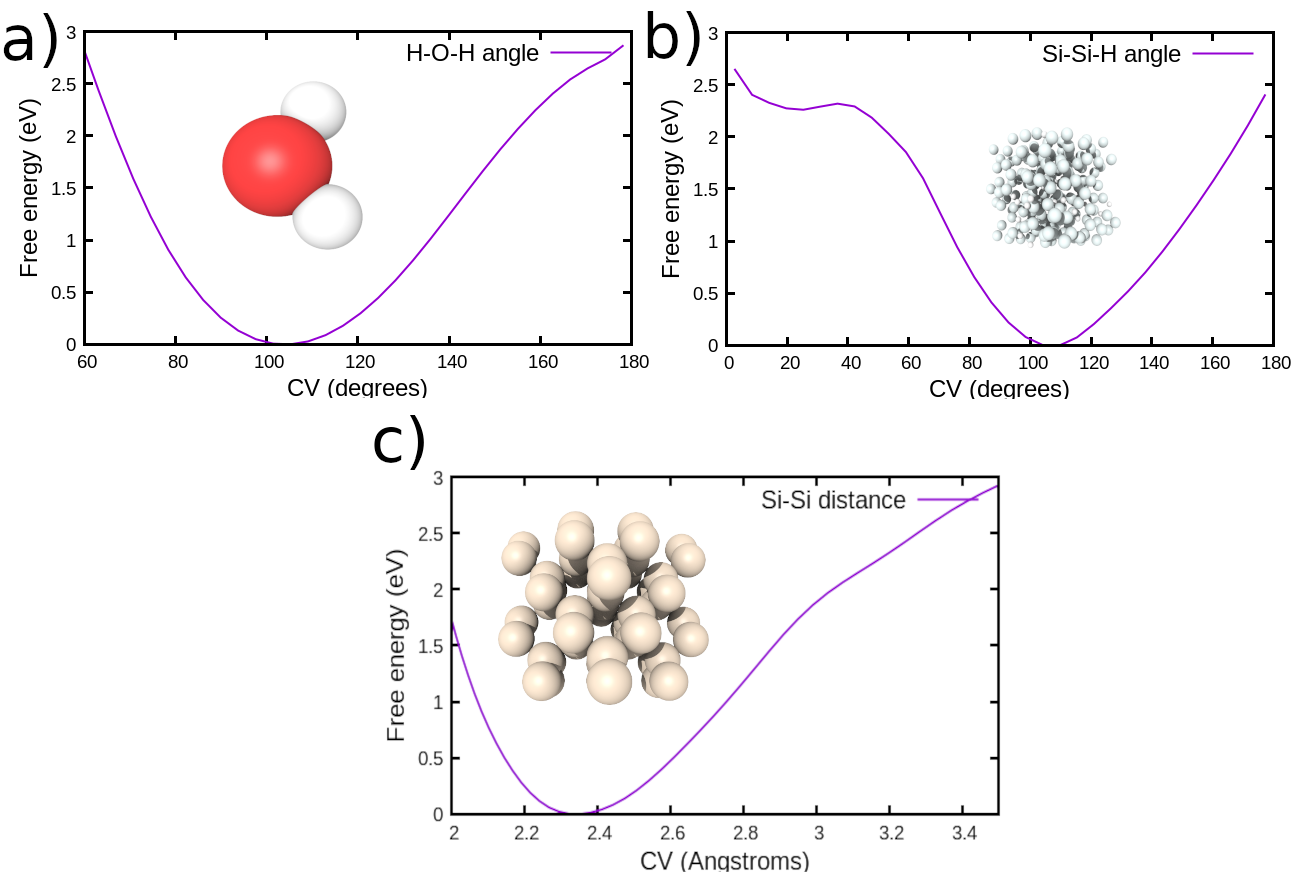}
    \caption{Free energy calculation of: a) Water internal angle from a \DeepMD model with \ac{ASE}, b) Si--Si--H angle of \ac{GAP} model with \ac{ASE} and c) Si--Si distance of a \ac{GNN} model with \JAXMD.}
    \label{fig:machine1}
\end{figure}

To test the capabilities of \PySAGES to handle different \ac{NN} force fields, 
we have selected three different systems trained with the methods mentioned above. 
For \DeepMD, we use a pre-trained model for water, where the enhanced sampling system is one single
water molecule in vacuum, the collective variable is the internal angle of the molecule and the sampling method is \ac{ABF} (\autoref{sec:abf}).
The results in \autoref{fig:machine1} show that the minimum for this free energy profile  is around 105 degrees, which is within the range of the experimental value. 

Next, in \autoref{fig:machine1}b, a \ac{GAP} potential was used for Si--H amorphous mixtures\;\cite{gap}. In this case, a system of 244 atoms was used, and the collective variable is the bond angle between a triad of Si--Si--H atoms in the mixture. 
The global minimum in free energy agrees with the histogram taken from unbiased simulations reported in\;\cite{gap}. 

Lastly, we studied a \ac{GNN} model of a Si crystal\;\cite{graph} with \PySAGES and \JAXMD.
In this case, a crystalline Si system of 64 atoms was used, and the \ac{CV} was the Si--Si distance for the the crystal.
The results of \autoref{fig:machine1}c show that for this model, the minimum in the free energy corresponds almost exactly to the experimental value for the Si--Si nearest distance of 2.35~\AA.

\subsection{Performance}\label{sec:performance}

Our analysis revealed that \PySAGES is at least $\sim$14--15 times faster than \ac{SSAGES} on an Nvidia \caps{V100} \ac{GPU} machine.
To obtain this estimate, we ran enhanced sampling using umbrella sampling along the center of mass distance between two spherical polymer domains to measure the free energy landscape of the fission of a spherical diblock-copolymer blend (\autoref{fig:polymer}) described in \autoref{sec:polymer}. For support and compatibility across libraries and \ac{MD} engine versions, we estimated the performance with \ac{SSAGES} \version{0.9.2}-alpha and \PySAGES \version{0.3.0} using \HOOMD \version{2.6.0} and \HOOMD \version{2.9.7}, respectively.


\subsubsection{\Ac{GPU} utilization analysis}\label{sec:gpu-performance}

\PPySAGES is designed to execute every compute-intensive step of a simulation on the \ac{GPU} and have zero copy instruction between \ac{GPU} device and host \ac{CPU} memory for its explicit backends for \HOOMD\;\cite{HOOMD-blue} and \OpenMM\;\cite{OpenMM}, while still providing Python code for the user through \JAX\;\cite{frostig2018compiling}.
In this section, we investigate the calculation efficiency of \PySAGES by examining two example systems, one for each backend.

For \HOOMD, we are investigating a system of highly coarse-grained \ac{DPD} diblock-copolymers as discussed in \autoref{sec:polymer}.
The simulation box contains a total of $nN=51\,200$ particles at a density of $\rho=51.2/\sigma^3$, which we use for benchmarking purposes with an Nvidia \caps{V100} \ac{GPU} hosted on an Intel Xeon Gold \caps{6248R} \ac{CPU} @~3.00GHz.
Running only with \HOOMD \version{2.9.7} we achieve an average \ac{TPS} of $754$, which is the expected high performance of \HOOMD on \acp{GPU}.

\begin{figure}
    \centering
    \includegraphics[width=\textwidth]{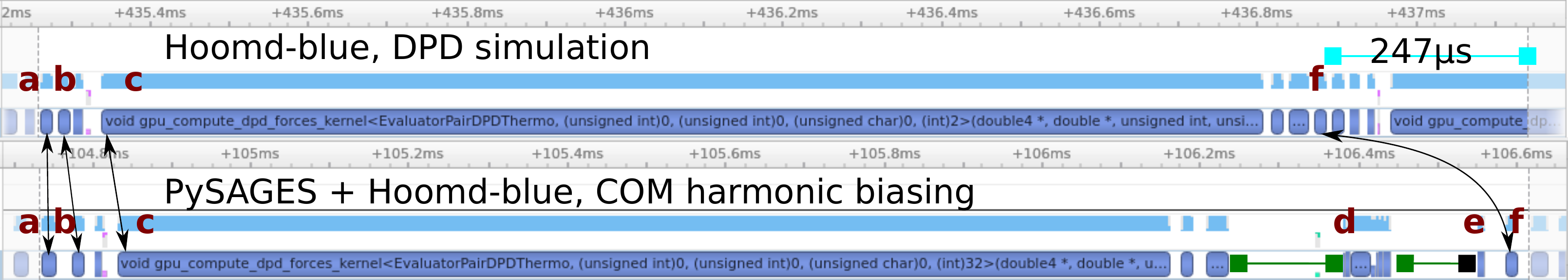}
    \caption{The figure shows a $1.8$ms section of profiled timeline recorded with Nvidia Nsight systems on an Nvidia \caps{V100} \ac{GPU}. The top row shows a vanilla \HOOMD simulation step, while the bottom row shows a \PySAGES/\HOOMD simulation with harmonic biasing of a center of mass \ac{CV}. Light-blue represents the \ac{GPU} activity while dark-blue represents individual \caps{CUDA} compute kernels. The maroon letters show case the same compute steps in both simulations: a) First half-step of integration, b) compute of bond forces, c) pair-forces, d) calculation of the \ac{CV}, e) addition of the harmonic biasing force to the \HOOMD simulation, and f) the second integration step. Sections d) and e) are \PySAGES only and are executed on the \ac{GPU}. We observe \ac{GPU} idle time during the \PySAGES Python coordination with \ac{GPU}--\JAX/\CuPy (green bar), but note that there is no memory copies even within the \ac{GPU} memory. The additional time for \ac{CV} biasing per time step is $247~\upmu$s (teal bar).}
    \label{fig:profile-hoomd}
\end{figure}

\autoref{fig:profile-hoomd} shows a detailed profiled timeline during the execution of a single time step.
During $1.8$ ms, \HOOMD spends the most computational effort on the calculation of pairwise \ac{DPD} forces.
It can be noted that \HOOMD is designed to have almost no idle time of the \ac{GPU} during a time step.
As soon as \PySAGES is added computation part, we observe that an additional part is added to calculate the \ac{CV} and add the forces to every particle.
This causes a small period of idle of the \ac{GPU}, since the execution also requires action of the Python runtime interface with \JAX.
In the future, we plan to launch the calculation of \ac{CV} asynchronously with the regular force calculation, which would hide this small \ac{CPU}-intensive \ac{GPU} idle time.
However, we measure that the total delay due to the extra computation is only about $247~\upmu$s only. We regard this to be an acceptable overhead for the user-friendly definition of \acp{CV}.

In order to connect multiple points in \ac{CV} space we can use enhanced sampling methods such as umbrella sampling (see \autoref{sec:umbrella-integration}) or the improved string method (see \autoref{sec:string}) to calculate the \ac{MFEP}.
Common for these advanced sampling methods that multiple replica of the system are simulations.
With \PySAGES we easily parallelize their execution using the Python module \texttt{mpi4py} and its \texttt{MPIPoolExecutor}.
This enables us to execute replica of the simulations on multiple \acp{GPU} even as they span different host machines.
In our example, we used 14 replicas for umbrella integration with 7 Nvidia \caps{V100} \acp{GPU}.
The use of a single \caps{V100} \ac{GPU} to execute the simulations with $5\cdot 10^5$ time steps for all replicas takes $2$ hours and $59$ minutes.
Ideal scaling with 7 \acp{GPU} reduces the time to solution to about $26$ minutes.
With our \ac{MPI}-parallel implementation, we achieve a time-to-solution of $28$ minutes.
Synchronization overhead and nonparallel aspects like final analysis sum up to $2$ minutes or about $9\%$ overhead.
This multi-\ac{GPU} implementation via \ac{MPI} enables automatically efficient enhanced sampling in \ac{HPC} environments.

For enhanced sampling methods that are designed for single replica simulations, we offer an implementation that allows multiple replicas to run in parallel, known as embarrassingly parallel computing.
In this situation, the build-in analysis averages the results from multiple replicas and estimates uncertainties.

\begin{figure}
    \centering
    \includegraphics[width=\textwidth]{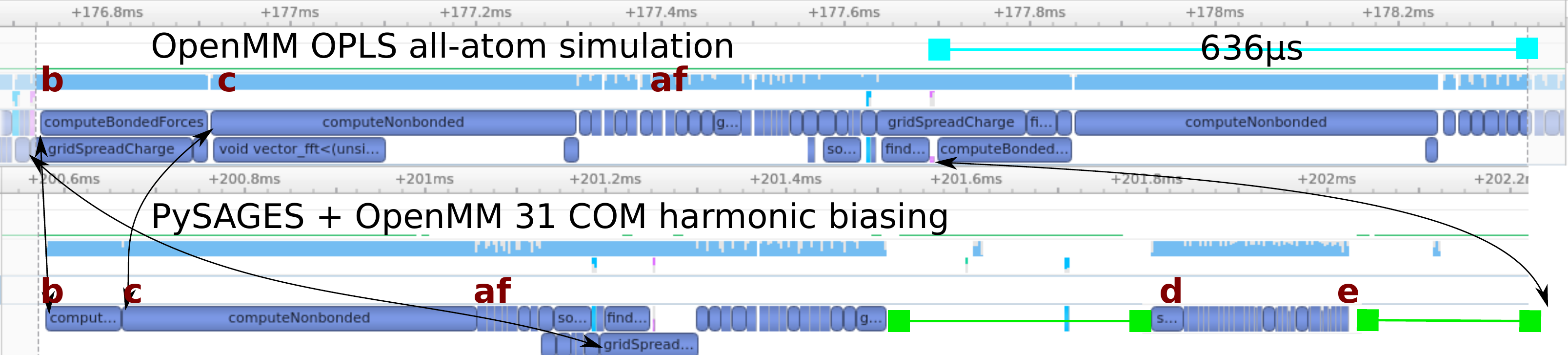}
    \caption{$1.6$ms profiled time line of an \OpenMM \caps{OPLS} simulation of $40,981$ particles as polymers with \ac{PME} summation for long-range Coulomb forces. The colors and labels are identical to \autoref{fig:profile-hoomd}. \OpenMM works with asynchronous \ac{GPU} kernel execution, which leads to less linearly sorted timelines, compared with \HOOMD, but we can still identify the \ac{CV} calculation d) and force biasing e) and the synchronization idle of the \ac{GPU} (green). Overall, the performance degradation is more pronounced with \OpenMM compared to \HOOMD.} 
    \label{fig:profile-openmm}
\end{figure}

In the previous section, we have demonstrated the fast \ac{GPU} interoperability between \PySAGES and \HOOMD via \JAX.
However, the concept of \PySAGES is to develop enhanced sampling methods independently of the simulation backend, so here we demonstrate that similar performance can be achieved with \OpenMM.
Since \OpenMM focuses on all-atom simulations, we simulate an all-atom model of a polymer
with the \BigSMILES\;\cite{lin2019bigsmiles} notation 
\texttt{\{[\$]CC([\$])(C)C(OCC(O)CSC1=CC=C(F)C(F)=C1)=O\}} 
with an \caps{OPLS}-\caps{AA} force field\;\cite{jorgensen1996development,schneider2022in} 
including long-range Coulomb forces via \ac{PME}.
We simulate a bulk system of 40mers with 31 macromolecules present, adding up to $40\,981$ atoms.
As a proof of concept, we calculated the center of mass for every polymer chain and biased it harmonically via \PySAGES.
As a performance metric, we evaluate the \ac{ns/day} executed on the same hardware configuration as a the \HOOMD example above.
For the unbiased, pure \OpenMM simulation we achieve a performance of $\approx136$ \ac{ns/day}.
For the \PySAGES biased simulation, we achieve a performance of $\approx 75$ \ac{ns/day}, equating to a biasing overhead of approximately $50\%$.
\autoref{fig:profile-openmm} shows a similar time series analysis as for \HOOMD.

It is notable that \OpenMM's execution model makes more use of parallel execution of independent kernels, which also changes the order of execution compared to \HOOMD.
As a result, the same \ac{CPU} synchronization changes the execution more drastically than in \HOOMD.
Additionally, a single time step for this system is faster executed compared to \HOOMD, making the synchronization overhead more noticeable.
In this case, parallelization of \PySAGES and \OpenMM is projected to have a bigger performance advantage.
Furthermore, we notice that the calculation of the center of mass and the biasing of all 31 polymer chains is more costly than the single \ac{CV} in the previous example.
The combination of these factors explain the higher \PySAGES overhead for this \OpenMM simulation, but overall performance is good and significantly better for alternative implementations that require \ac{CV} calculations on the \ac{CPU}.

\section{Conclusion}\label{sec:conclusion}

We have introduced \PySAGES, a library for enhanced sampling
in molecular dynamics simulations, which allows users to
utilize a variety of enhanced sampling methods and
collective variables, as well as to implement new ones via a
simple Python and \JAX-based interface.

We showed how \PySAGES can be used through a number of example
applications in different fields such as drug design,
materials engineering, polymer physics, and ab-initio MD
simulations. We hope that these convey for the reader the
flexibility and potential of the library for addressing a
diverse set of problems in a high-performance manner.

As our analysis showcased, for large problems, \PySAGES can
perform biased simulation well over one order of magnitude faster than a library such as
\ac{SSAGES} even when the backend already performs
computations on a \ac{GPU}.

Nevertheless, as with any newly developed software,
\PySAGES is still under development and we are continually working to improve it.
In the near term, we plan to add the ability for users to perform restarts,
which will provide greater flexibility running long simulations.
Moreover, we plan to optimize \PySAGES-side computations to run fully
asynchronously with the computation of the forces of the backend,
which will further enhance its current performance.
We also invite the community to contribute to the development of \PySAGES,
whether by suggesting new features, reporting bugs, or contributing code.

Overall, we believe that \PySAGES provides a useful tool
for researchers interested in performing molecular and ab-initio
simulations in multiple fields, due to its user-friendly 
framework for defining and using sampling methods and
collective variables, as well as its high performance on
\ac{GPU} devices.



Looking further ahead, we are excited about the potential for
\PySAGES to enable fully end-to-end differentiable free energy calculations.
This will provide new possibilities for force-field and materials design,
which would drive significant advances in these areas.

\section*{Code avalability}
The code for \PySAGES is available in the GitHub repository: \url{https://github.com/SSAGESLabs/PySAGES}.

\section*{Acknowledgements}
This work is supported by the Department of Energy, Basic Energy Sciences, Materials Science and Engineering Division, through the Midwest Integrated Center for Computational Materials (\caps{MICC}o\caps{M}).
{L.\,S.} is grateful for the support of the Eric and Wendy Schmidt AI in Science Postdoctoral Fellowship at the University of Chicago.
{R.\,A.} is supported by the Dutch Research Council (\caps{NWO} Rubicon \caps{019.202EN.028}).
The authors also acknowledge the Research Computing Center of the University of Chicago for computational resources.

\section*{Conflict of Interest Statement}

{A.\,L.\,F.} is a co-founder and consultant of Evozyne, Inc. and a co-author of US Patent Applications \textsc{16/887,710} and \textsc{17/642,582}, US Provisional Patent Applications \textsc{62/853,919}, \textsc{62/900,420}, \textsc{63/314,898}, and \textsc{63/479,378} and International Patent Applications \caps{PCT/US2020/035206} and \caps{PCT/US2020/050466}.

\newpage 

\appendix

\section{\Acl{CV} for the distance to an interface}%
\label{appendix:interface}
Implementation of the \ac{CV} described in
\autoref{sec:cvs-interface}, that is, the distance between a
group of atoms to an interface defined by another group of
atoms.

\begin{minted}[
  baselinestretch=1.25,
  bgcolor=black!2,
  escapeinside=||,
  fontsize=\scriptsize,
  xleftmargin=1em,
]{python}
||
class DistanceToInterface(TwoPointCV):
    def __init__(self, indices, axis, sigma, scope, bins=100, coeff=1):
        super().__init__(indices)
        self.axis = axis
        self.sigma = sigma
        self.scope = scope
        self.bins = bins
        self.coeff = coeff

    @property
    def function(self):
        return lambda r1, r2: distance_to_interface(
            r1, r2, axis=self.axis,
            sigma=self.sigma, scope=self.scope,
            bins=self.bins, coeff=self.coeff
        )

def distance_to_interface(p1, p2, axis, sigma, scope, bins, coeff):
    mobile_axis = barycenter(p1)[axis]
    positions_axis = p2.flatten()[axis::3]
    centers = np.linspace(scope[0], scope[1], bins)
    centers = np.expand_dims(centers, 1)
    positions_axis = np.expand_dims(positions_axis, 0)
    diff = positions_axis - centers
    mass = np.exp(-0.5 * (diff / sigma) ** 2)
    mass = np.sum(mass, axis=1)
    mass_diff = np.abs(mass[1:] - mass[:-1])
    centers = np.squeeze(centers)
    centers_mean = (centers[1:] + centers[:-1]) / 2
    probability = nn.softmax(mass_diff * coeff)
    interface = np.sum(probability * centers_mean)
    return mobile_axis - interface
||
\end{minted}

\newpage 

\section{Benchmark test systems}\label{appendix:benchmark-systems}

In the following sections, we present the results of the
free energy calculation for the benchmark test systems of
\acl{ADP} and butane. The details of all the parameters
chosen to perform the enhanced sampling simulation of these
are summarized in \ref{appendix:details}.

\subsection{Alanine Dipeptide}\label{appendix:alanine}
\begin{figure}[htbp]
  \centering
  \includegraphics[width=\textwidth]{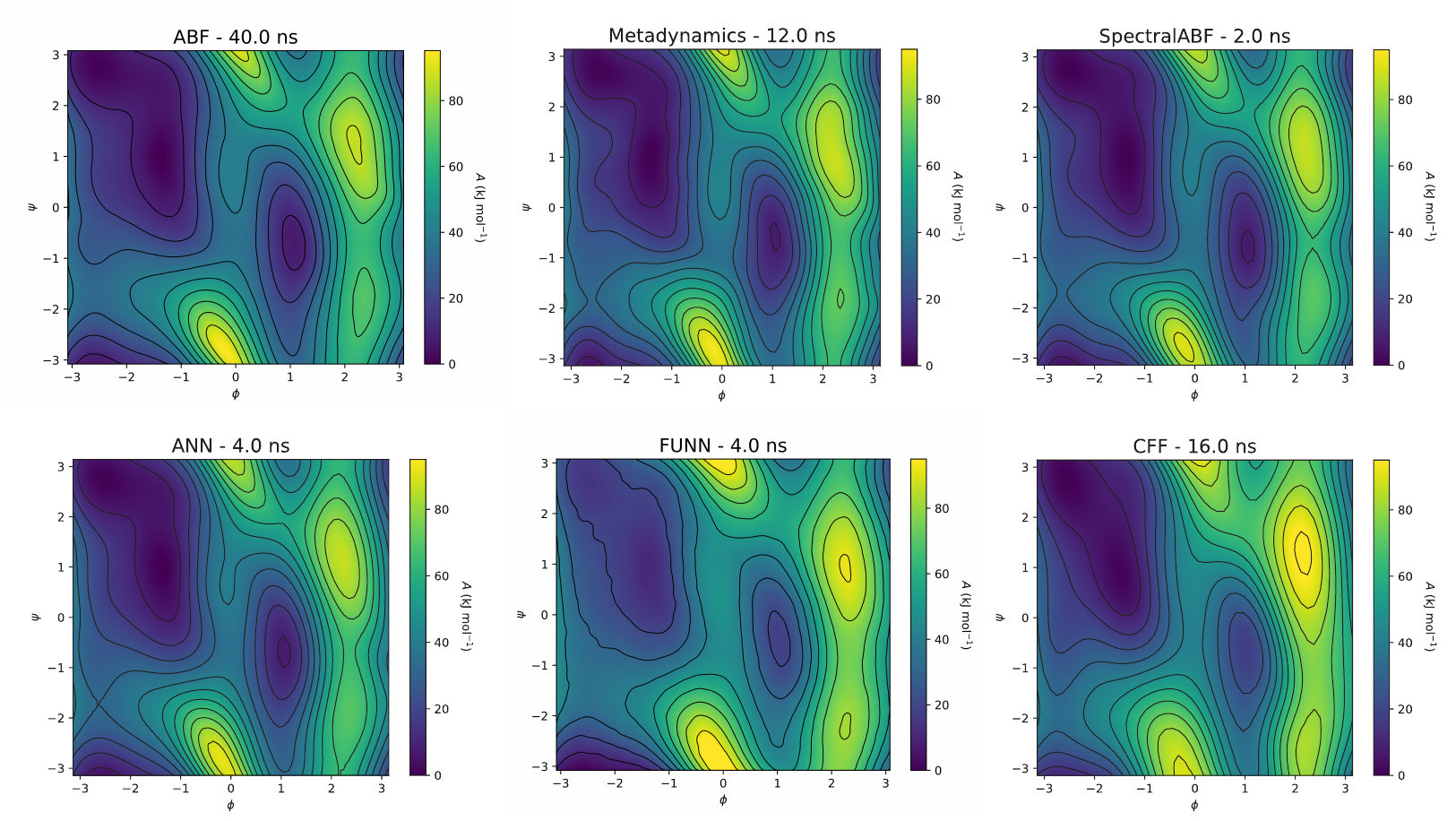}
  \caption{
    Free energy landscape of alanine dipeptide (Amber f\,f\caps{99sb}\;\cite{amber99sb})
    in vacuum as a function of the dihedral angles $\phi$ and $\psi$ obtained 
    with \PySAGES and \OpenMM via different enhanced sampling methods: 
    \ac{ABF}, Metadynamics, Spectral~\ac{ABF}, \ac{ANN}, \ac{FUNN}, \ac{CFF}.
    Each panel also indicates the length of the simulation necessary for the
    free energy to converge.
    The long \ac{ABF} simulations represent the ground truth.
  }
  \label{fig:adp}
\end{figure}

The first test system involves \acl{ADP} in vacuum (Figure~\ref{fig:adp}), a
benchmark system for enhanced sampling methods that is frequently used in the literature.

\subsection{Butane}\label{appendix:butane}
\begin{figure}[htbp]
    \centering
    \includegraphics[width=0.5\textwidth]{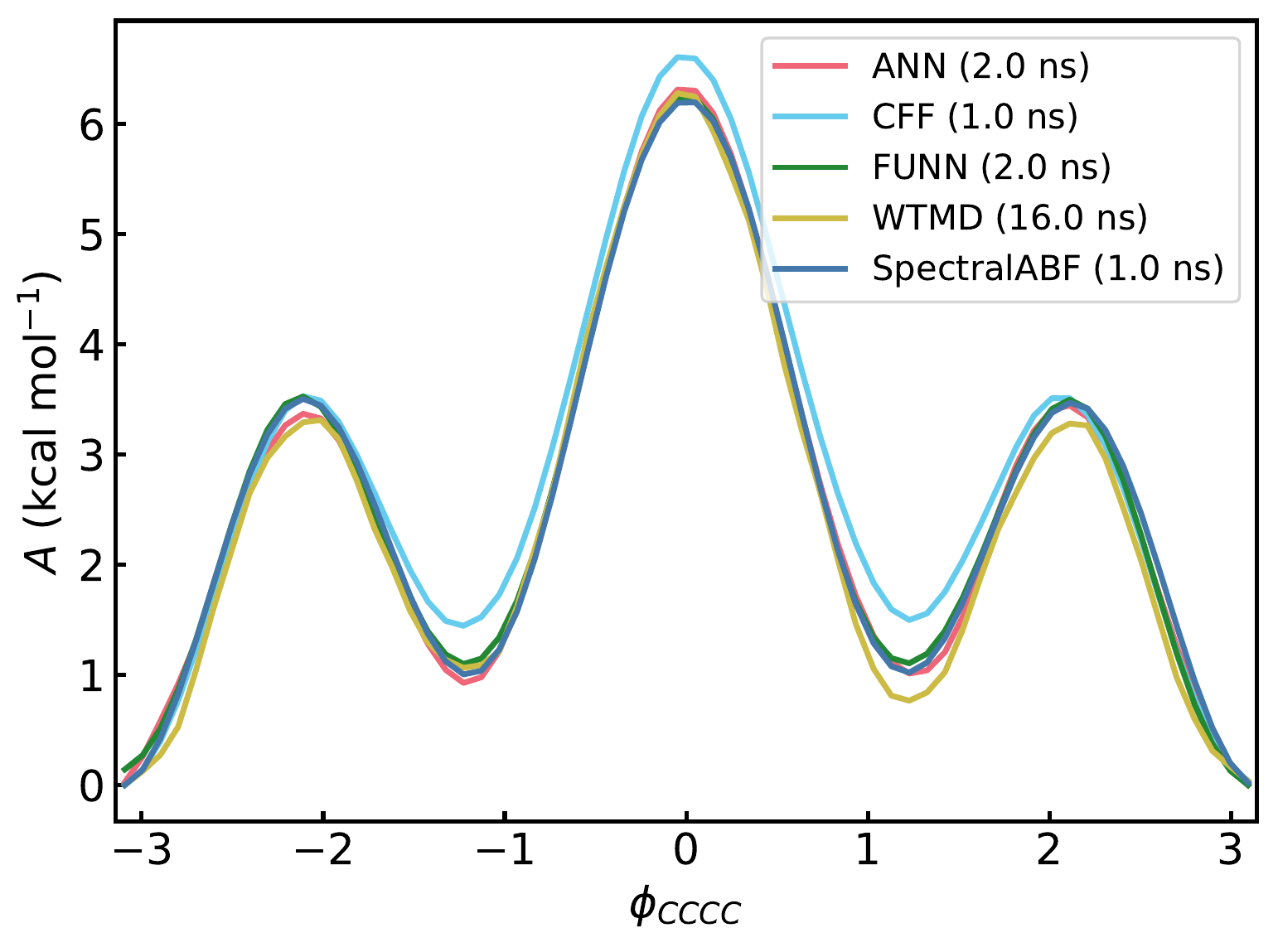}
    \caption{Free energy profile along the dihedral angle of a butane molecule
    (using an \caps{OPLS}-based force field\;\cite{jorgensen1996development})
    obtained via different enhanced sampling methods with \PySAGES and \HOOMD:
    \ac{ANN}, 
    \ac{CFF}, 
    \ac{FUNN}, Spectral~\ac{ABF}, \ac{WTMD}.
    The legend also indicates the length of the simulation.
    The long \ac{ABF} simulations represent the ground truth.
    }
    \label{fig:butane}
\end{figure}

As a second test system, we compute the free energy profile along the C-C-C-C 
dihedral angle, $\phi_{CCCC}$, of a butane molecule (in vacuum), Figure~\ref{fig:butane}.

\subsection{Example System Details}\label{appendix:details}

\begin{table}[htbp]
\label{tbl:parameters}
\centering
\caption{
    Parameters and methods details for the various examples. For all methods but Metadynamics,
    we used a grid with 50 points along each \ac{CV} for \ac{ADP} and with 64 points along
    the \ac{CV} for butane.\\
    $N$ (\ac{ABF}) = Threshold parameter before accounting for the full average
    of the adaptive biasing force.\\
    \ac{ADP} = \acl{ADP}
} 
\vspace{2mm} 
\smaller
\begin{tabular}{llclll}
\hline\\[-2ex]
\textbf{System}&\textbf{Backend}&\textbf{\ac{CV}}&\textbf{Method}&\textbf{Settings}&\textbf{Fig.}\\
\\[-2ex]\hline\\[-2ex]
\multirow{8}{*}{\ac{ADP}}   &\multirow{8}{*}{\OpenMM}&\multirow{8}{*}{$\phi$ and $\psi$} 
                       & \ac{ABF}       &$N=500$ (default)&\multirow{8}{*}{\ref{fig:adp}}\\
                       &                       &                       
                       & \ac{ANN}       &$\mathrm{topology}=(8,8)$ &    \\
                       &                       &                       
                       & \ac{CFF}       &$\mathrm{topology}=(14,)$ &    \\
                       &                       &                       
                       & \ac{FUNN}      &$\mathrm{topology}=(14,)$ &    \\
                       &                       &                       
                       &\multirow{3}{*}{Metadynamics}&$\sigma=0.35$ rad &    \\
                       &                       &                       
                       &                             & $h=1.2$ kJ/mol  &    \\
                       &                       &                       
                       &                             & stride $= 500$  &    \\        
                       &                       &                       
                       & Spectral~\ac{ABF}    & ---                      &    \\
\\[-2ex]\hline\\[-2ex]
\multirow{8}{*}{Butane}&\multirow{8}{*}{\HOOMD}&\multirow{8}{*}{$\phi_{CCCC}$} 
                       & \ac{ANN}   &$\mathrm{topology}=(8,8)$  &\multirow{8}{*}{\ref{fig:butane}}\\
                       &                       &                       
                       & \ac{CFF}   &$\mathrm{topology}=(8,)$   &    \\
                       &                       &                       
                       & \ac{FUNN}  &$\mathrm{topology}=(8,)$   &    \\
                       &                       &                       
                       &\multirow{3}{*}{\ac{WTMD}}&$\sigma=0.10$ rad &    \\
                       &                       &                       
                       &                             & $h=0.01$ kJ/mol  &    \\
                       &                       &                       
                       &                             & stride $= 50$  &    \\
                       &                       &                       
                       &                             & $\Delta T= 5000$ &    \\
                       &                       &                       
                       & Spectral~\ac{ABF}& ---                       &    \\
\\[-2ex]\hline
\end{tabular}
\end{table}

\newpage 

\bibliographystyle{elsarticle-num} 
\bibliography{references}

\end{document}